\documentclass[a4paper,12pt, epsfig,longbibliography,nofootinbib]{article}
\def\letter{0}\def\pr{0}
\pdfoutput=1 
\usepackage{epsfig}
\usepackage{epstopdf}
\usepackage{graphicx}
\usepackage{ifthen}
\usepackage[square,numbers,sort&compress]{natbib}
\usepackage{braket}
\usepackage{slashed}
\usepackage{cancel}

\usepackage{amsmath}
\usepackage[psamsfonts]{amssymb}
\usepackage{euscript}

\usepackage{latexsym}
\usepackage[arrow,matrix,curve]{xy}

\usepackage[hypertexnames=false]{hyperref}
\hypersetup{
    colorlinks=true,
    linkcolor=blue,
    filecolor=magenta,   
    citecolor=black,   
    urlcolor=cyan,
    }

\newskip\humongous \humongous=0pt plus 1000pt minus 1000pt

\newif\ifdtup

\def\,{\hspace{-.1cm}}
\def\hsp{,\hspace{.7cm}}

\def\fc#1#2 {\frac{n}{q}#1\frac{n}{q}#2}

\newcommand{\vac}{\ensuremath{|0\rangle}}

\renewcommand{\tanh}{\textrm{tanh}}
\newcommand{\sech}{\textrm{sech}}
\newcommand{\csch}{\textrm{csch}}

\def\exp#1{\hbox{\rm exp}\left[#1\right]}

\newcommand{\intdk}{\int\frac{dk}{2\pi}}

\renewcommand{\theequation}{\arabic{section}.\arabic{equation}}
\renewcommand{\(}{\begin{equation}}
\renewcommand{\)}{end{equation} \vspace{-.05in}\linebreak}

\newcounter{saveeqn}
\newcounter{savealpheqn}

\newcommand{\alpheqn}{\setcounter{saveeqn}{\value{equation}}%
  \stepcounter{saveeqn}\setcounter{equation}{0}%
  \renewcommand{\theequation}{\mbox{\arabic{section}.\arabic{saveeqn}
\alph{equation}}}
  \renewcommand{\)}{\end{equation}}}
\def\part#1{\frac{\partial}{\partial{#1}}}%
\def\group#1{\refstepcounter{equation}\setcounter{saveeqn}
 {\value{equation}}%
  \label{#1}\setcounter{equation}{0}%
\renewcommand{\theequation}{\mbox{\arabic{section}.\arabic{saveeqn}
\alph{equation}}}
  \renewcommand{\)}{\end{equation}}}
\newcommand{\reseteqn}{\setcounter{equation}{\value{saveeqn}}%
  \renewcommand{\theequation}{\arabic{section}.\arabic{equation}}%
  \renewcommand{\)}{\end{equation}}}

\newcommand{\aalpheqn}{\setcounter{saveeqn}{\value{equation}}%
  \stepcounter{saveeqn}\setcounter{equation}{0}%
  \renewcommand{\theequation}{\mbox{
        \Alph{subsection}.\arabic{saveeqn}\alph{equation}}}
   \renewcommand{\)}{\end{equation}}}
\newcommand{\areseteqn}{\setcounter{equation}{\value{saveeqn}}%
  \renewcommand{\theequation}{\Alph{subsection}.\arabic{equation}}%
  \renewcommand{\)}{\end{equation}}}

\renewcommand{\thefootnote}{\alph{footnote}}
\renewcommand{\(}{\begin{equation}}
\renewcommand{\)}{\end{equation}}
\newcommand{\ba}{\begin{eqnarray}}
\newcommand{\ea}{\end{eqnarray}}

\renewcommand{\sl}{{\sqrt{\lambda}}}

\newcommand{\cbp}{\mathop{\vtop{\ialign{##\crcr
   $\hfil\displaystyle{}\hfil$\crcr\noalign{\kern-13pt\nointerlineskip}
   \BIG{)}\hskip0pt\crcr\noalign{\kern3pt}}}}}
\newcommand{\pa}{\mathop{\vtop{\ialign{##\crcr

$\hfil\displaystyle{\oplus}\hfil$\crcr\noalign{\kern+1pt\nointerlineskip
}
   \hspace{.08in}$^{\alpha=0}$\hskip6pt\crcr\noalign{\kern3pt}}}}}
\renewcommand{\hsp}{,\hspace{.3in}}
\newcommand{\p}{^\prime}

\catcode`\@=11
\def\vereq#1#2{\lower3pt\vbox{\baselineskip1.5pt \lineskip1.5pt
\ialign{$\m@th#1\hfill##\hfil$\crcr#2\crcr\sim\crcr}}}
\catcode`\@=12

\renewcommand{\(}{\begin{equation}}
\renewcommand{\)}{\end{equation}}

\def\pin#1{\int \frac{d#1}{2\pi}}

\def\df{\mathcal{D}_{f}}

\def\os{\omega_S}

\newcommand{\beas}{\begin{eqnarray*}}
\newcommand{\eeas}{\end{eqnarray*}}

\newcommand{\bquo}{\begin{quote}}
\newcommand{\enqu}{\end{quote}}

\def\lim#1{\stackrel{\rm{lim}}{{}_{#1}}}

\newcommand{\g}{\mathfrak g}

\def\ok#1{\omega_{k_{#1}}}
\def\okp#1{\omega_{k\p_{#1}}}

\def\V#1{V^{(#1)}(\sqrt{\lambda}f(x))}

\newcommand{\beq}{\begin{equation}}
\newcommand{\eeq}{\end{equation}}
\newcommand{\bea}{\begin{eqnarray}}
\newcommand{\eea}{\end{eqnarray}}
\newcommand{\bal}{\begin{align}}
\newcommand{\eal}{\end{align}}

\newskip\humongous \humongous=0pt plus 1000pt minus 1000pt

\newif\ifdtup

\catcode`\@=11

\ifthenelse{\equal{\letter}{0}}{ 

\@addtoreset{equation}{section}
\def\theequation{\arabic{section}.\arabic{equation}}

\def\@normalsize{\@setsize\normalsize{15pt}\xiipt\@xiipt
\abovedisplayskip 14pt plus3pt minus3pt%
\belowdisplayskip \abovedisplayskip
\abovedisplayshortskip \z@ plus3pt%
\belowdisplayshortskip 7pt plus3.5pt minus0pt}

\def\small{\@setsize\small{13.6pt}\xipt\@xipt
\abovedisplayskip 13pt plus3pt minus3pt%
\belowdisplayskip \abovedisplayskip
\abovedisplayshortskip \z@ plus3pt%
\belowdisplayshortskip 7pt plus3.5pt minus0pt
\def\@listi{\parsep 4.5pt plus 2pt minus 1pt
      \itemsep \parsep
      \topsep 9pt plus 3pt minus 3pt}}

\relax

\def\section{\@startsection{section}{1}{\z@}{3.5ex plus 1ex minus  .2ex}{2.3ex plus .2ex}{\large\bf}}

\def\thesection{\arabic{section}}
\def\thesubsection{\arabic{section}.\arabic{subsection}}

\def\appendix{\setcounter{section}{0}
 \def\thesection{Appendix \Alph{section}}
 \def\thesubsection{\Alph{section}.\arabic{subsection}}
 \def\theequation{\Alph{section}.\arabic{equation}}}
\renewcommand{\theequation}{\arabic{section}.\arabic{equation}}

}{
\renewcommand{\theequation}{\arabic{equation}}

} 

\begin{document}
\def\thefootnote{\fnsymbol{footnote}}

\def\thetitle{A Resonance in Elastic Kink-Meson Scattering}
\def\autone{Bilguun Bayarsaikhan}
\def\auttwo{Jarah Evslin}

\def\affa{Institute of Modern Physics, NanChangLu 509, Lanzhou 730000, China}
\def\affb{University of the Chinese Academy of Sciences, YuQuanLu 19A, Beijing 100049, China}

\ifthenelse{\equal{\pr}{1}}{
\title{\thetitle}
\author{\autone}
\author{\auttwo}
\author{\autthree}
\affiliation {\affa}
\affiliation {\affb}
\affiliation {\affc}

}{

\begin{center}
{\large {\bf \thetitle}}

\bigskip

\bigskip


{\large \noindent  \autone{${}^{12}$}
\footnote{ph.bilguun@gmail.com}
and  
\auttwo{${}^{12}$}
\footnote{jarah@impcas.ac.cn}
}

\vskip.7cm

1) \affa\\
2) \affb\\

\end{center}

}

\begin{abstract}
\noindent
We analytically sum the leading bubble diagrams that contribute to the elastic scattering amplitude of a kink and a meson in the $\phi^4$ double-well model.  We find a single peak, corresponding to the unstable kink state in which the shape mode is excited twice.  The peak has the usual Breit-Wigner form, and its imaginary part agrees with the shape mode decay rate found by Manton and Merabet.

\end{abstract}

%

\setcounter{footnote}{0}
\renewcommand{\thefootnote}{\arabic{footnote}}

\ifthenelse{\equal{\pr}{1}}
{
\maketitle
}{}

\section{Introduction}
\label{sec:intro}

\subsection{Unstable Excitations of Solitons}

Calculating the spectrum of perturbations of solitons is a major industry \cite{solm1,sol0,sol1,sol2,sol3}, and no paper about a new soliton solution is complete without an analysis of its perturbations.  The perturbations determine whether the soliton is perturbatively stable, they yield its spectrum of excitations, and they even affect its scattering with other solitons \cite{res1,res2,res3,res4,res5,res6}.  

The calculations proceed as follows.  One inserts a small perturbation of the soliton solution into the equations of motion.  The solutions, at linear order in the perturbations, are the linearized perturbations.  For a stable soliton, these will be stable.  At nonlinear order, the situation is quite different and such perturbations generally \cite{resconj1,resconj2} decay to radiation.  One example is the shape mode of the kink in the classical $\phi^4$ double-well model, whose decay rate was calculated in Ref.~\cite{mm}.

In quantum field theory, each such linearized perturbation corresponds to a mode which may be excited.  The classical nonlinear instability is lifted to the statement in quantum field theory that the mode, when multiply excited\footnote{In quantum field theory, the nonlinearity of the decay rate arises from the fact that multiple excitations of the bound mode are needed to create one quantum of radiation.}, lies above the mass gap and so decays into bulk excitations, which we will call mesons.  For example, in the case of the $\phi^4$ model, the shape mode is stable if excited once, but if excited twice its energy lies above the mass gap and its lifetime was calculated in Ref.~\cite{Evslin_2022}.  That lifetime for a single pair of excitations, when applied to a coherent state, reproduced the decay rate in classical field theory of Ref.~\cite{mm}.

The existence of such unstable excitations of quantum solitons is therefore quite generic.  For example, if $B^\dag$ is an operator which excites a stable, bound perturbation, then some power of $B^\dag$ generally excites an unstable perturbation.  But, in general, how can the spectra of unstable perturbations be found?  {\it{The present paper will describe a novel way to find the energies and lifetimes of unstable excitations of solitons.}}  The strategy will be to consider the scattering of a soliton with a meson.  The locations of the peaks in the elastic scattering amplitude will yield the energies of the unstable soliton excitations, while the widths of the peaks yield the lifetimes.   Of course in particle physics this is all very standard, but the fact that it extends to kinks is nontrivial because, for example, the meson is a perturbation of a vacuum but the vacua themselves on the two sides of the kink are different.

\subsection{Decay Rates}

The lifetime of an unstable perturbation in quantum field theory can be used to compute the corresponding radiation rate in the classical field theory.  Of course this correspondence between classical and quantum decay rates only holds at tree level.  In quantum field theory, one expects loop corrections to the decay rate.  Can they be calculated?

It has long been appreciated \cite{Levy_1959} that the lifetime of an unstable excitation is, beyond the leading orders in perturbation theory, in general ill-defined because it depends on the arbitrary choice of initial state.  For this reason one is instead interested in the width of the resonance corresponding to the creation of the virtual excitation, which agrees with the inverse lifetime at leading orders in a certain adiabatic approximation \cite{Ram_1970}.  {\it{In the present note we will calculate, for the first time, the leading order width of a resonance corresponding to an excitation of a topological soliton, and we will show that it agrees with the known leading order inverse lifetime.}}

More specifically, we will consider a model of a scalar subjected to a potential with degenerate minima.  Such a model possesses kinks, unbound mesons, and in many cases also possesses shape modes, which are bound to the kinks
.  A shape mode is bound in the sense that its excitation energy $\omega_S$ is less than the mass gap of the theory.  However we will restrict our attention to models such that $2\omega_S$ is more than the mass gap, so that if the shape mode is excited with occupation number two, then it may escape into the continuum.  Although the generalization is rather trivial, for simplicity we will restrict our attention to classically reflectionless kinks such as those of the $\phi^4$ double-well model mentioned above.  

The resonance that we will study will be a peak in the elastic kink-meson scattering amplitude.  Classically, of course, there is no elastic scattering because we have restricted our attention to reflectionless kinks.  However, elastic scattering appears at the next order in the semiclassical expansion.  The corresponding amplitude was first considered in Ref.~\cite{Uehara_1991} using the collective coordinate approach of Ref.~\cite{Gervais_1975}.  However, due to the complexity of the collective coordinate method \cite{cch1,cch2}, contributions from intermediate states with two excitations were explicitly dropped, although they contribute at the same order as the processes that were considered.  It is precisely such an intermediate state which leads to the peak studied here.

Recently a simpler  framework to address such questions, Linearized Soliton Perturbation Theory
(LSPT), has appeared  \cite{Evslin_2019}.  It was used to address this problem at leading nonvanishing order in the case of a general potential in Refs.~\cite{evslin2023elastickinkmesonscattering,evslin2024reflectioncoefficientreflectionlesskink}.  In Ref.~\cite{Bilguun2025elastickinkmesonscatteringphi4} the authors restricted attention to the $\phi^4$ double-well model.  They found that the S-channel process containing an intermediate state with a twice-excited shape mode contributes the usual S-channel pole as the incoming meson energy approaches twice the shape mode energy,
$\omega_{k_0}\to 2\omega_S$, where $\omega_{k_0}$ is the energy of the incoming meson. Physically, this corresponds to virtually exciting a nearly on-shell
intermediate kink configuration in which the shape mode is excited twice.

As always, S-channel poles resulting from on-shell virtual particles are smeared out by higher order corrections.  The goal of this paper is to resum the leading bubble corrections that dress this twice-excited
shape mode intermediate state in the near-resonant regime. We find that, as is usual when dealing with particles, bubble diagrams smooth the tree-level pole into a Breit-Wigner resonance.

The paper is organized as follows. In Sec.~\ref{lsptsez} we review LSPT and fix our conventions. In Sec.~\ref{Sec3} we summarize the leading order elastic amplitude and isolate the
analytic structure associated with the intermediate state
\cite{Bilguun2025elastickinkmesonscatteringphi4}. In Sec.~\ref{Sec4} we derive the bubble chain directly
from the Schr\"odinger-picture Hamiltonian eigenvalue problem
and extract the pole shift and width, comparing with Ref.~\cite{Evslin_2022}.
In Sec.~\ref{modsez} we evaluate the width and line shape in the $\phi^4$ double-well
model. We conclude in Sec.~\ref{sec:remarks}.

\section{Linearized Soliton Perturbation Theory} \label{lsptsez}

In this section, we briefly review Linearized Soliton Perturbation Theory (LSPT) as formulated in Refs.~\cite{Evslin_2019,Evslin_2020}.  It is a formalism for lifting nonlinear solutions of classical field theory to states in quantum field theory, and then computing observables associated with those states, such as form factors, scattering amplitudes, excitation spectra and their lifetimes and more.

\subsection{The displacement operator}


In the present study, we will apply LSPT to a $(1+1)$-dimensional theory of a real scalar field $\phi(x)$ and its conjugate momentum $\pi(x)$ in the Schrödinger picture.  Our model will be defined by the Hamiltonian 
\begin{equation}
    H=\int d x: \mathcal{H}(x): ,
\end{equation}
where the standard Schrodinger-picture plane-wave normal ordering $::$ removes the tadpole divergences, which are the only ultraviolet divergences in the model.  The Hamiltonian density is
\begin{equation}
 \quad \mathcal{H}(x)=\frac{\pi^2(x)}{2}+\frac{\partial_x \phi(x)\partial_x \phi(x)}{2}+\frac{1}{\lambda} V(\sqrt{\lambda} \phi(x)).   
\end{equation}
We assume that the potential $V(\sqrt{\lambda} \phi(x))$ has two degenerate minima, so that the classical equations admit a static kink solution $\phi(x)=f(x)$ which interpolates between the minima. We define
\begin{equation}
V^{(n)}(\sqrt{\lambda} \phi(x))=\frac{\partial^n V(\sqrt{\lambda} \phi(x))}{\partial (\sqrt{\lambda} \phi(x))^n}
\end{equation}
and the (vacuum) meson mass by the asymptotic second derivative along the kink profile,
\begin{equation}
    m^2=V^{(2)}(\sqrt{\lambda} f(x))\Big|_{x=\pm\infty}\label{dec}.
\end{equation}
We assume the masses in the two vacua at $x=\pm\infty$ coincide so that the kink does not experience a one-loop acceleration \cite{Weigel_2019}. Choosing a particular static kink profile $f(x)$ singles out a definite position for the kink and thus makes translation invariance non-manifest in intermediate steps. The underlying theory, however, remains translation invariant.  In fact, translation invariance reappears order by order \cite{Evslin_2020}.

In the weak-coupling limit, kink states in the quantum theory do not belong to the usual vacuum sector Fock space built by acting finite numbers of mesonic plane-wave creation operators on the vacuum. Instead, they live in a distinct one-kink sector.  This is a Fock space which consists of the kink ground state, acted upon by a finite number of bound (shape) and continuum normal-mode (meson) creation operators, as well as an arbitrary function of the zero-mode operators.  This arbitrary function is fixed for a momentum eigenstate.  As this one-kink sector lies outside of the $n$-meson Fock space, it is not directly accessible within the ordinary vacuum sector perturbation theory.
Indeed, at weak coupling, in the vacuum sector the expectation value of the field $\phi(x)$ is suppressed by a power of $\sqrt{\hbar}$, and so vanishes in the classical limit, whereas in the kink sector the expectation value is approximately equal to $f(x)$, which is independent of $\hbar$.  
In the Sch\"odinger wave functional approach, one would say that the one-kink sector is spanned by the field eigenvectors whose eigenvalues are equal to $f(x)$ plus small perturbations.

In classical field theory, we would construct such small perturbations by decomposing the classical field $\phi(x,t)$ as
\beq
\phi(x,t)=f(x)+\g(x,t).
\eeq
Then we would define a kink Hamiltonian $H\p$ by
\begin{equation}
    H'[\g,\pi]\equiv H[\g+f,\pi]=H[\phi,\pi], \label{hp}
\end{equation}
which generates the same classical evolution and fluctuation spectrum, but has the practical advantage that configurations close to the kink can be treated perturbatively because the moments of $\g$ are small. 

But how does one apply the same idea in quantum field theory? The complication is that in the quantum field theory, the Hamiltonian must be regularized due to the ultraviolet divergences, and there is no general guarantee that a given regularization prescription is invariant under the shift $\phi\rightarrow\phi-f=\g$.  The simplest possibility is using the $H\p$ and $H$ defined in Eq.~(\ref{hp}) but then regularizing $H\p$ and $H$ with regulators $\Lambda\p$ and $\Lambda$ respectively, yielding $H\p_{\Lambda\p}$ and $H_\Lambda$.  The problem is that the regularized Hamiltonians are no longer equal $H\p_{\Lambda\p}[\g,\pi]\neq H_\Lambda[\phi,\pi]$, and so quantities such as the kink mass will depend on both $\Lambda\p$ and $\Lambda$.  More seriously, when $\Lambda\p$ and $\Lambda$ are both taken to infinity, it was shown in Ref.~\cite{Rebhan:1997iv} that the kink mass depends on the arbitrary choice of direction in the $(\Lambda\p,\Lambda)$ plane in which this limit is taken. This is to be expected: As the theory is defined by the original Hamiltonian $H_\Lambda$, its spectrum is the physical one, and masses computed using the naively shifted Hamiltonian $H\p_{\Lambda\p}$ correspond to a different theory. 

The displacement operator approach in the LSPT resolves this problem by defining the unitary displacement operator, $\df$
\begin{equation}
\df=\exp{-i\int dx f(x) \pi(x)},
\end{equation}
which implements the shift by the classical profile,
\begin{equation}
\mathcal{D}_{f}^{\dagger}\phi(x)\mathcal{D}_f=\phi(x)+f(x), \quad \mathcal{D}_{f}^{\dagger}:F[\phi(x),\pi(x)]:\df=:F[\phi(x)+f(x),\pi(x)]:
\end{equation}
for an arbitrary functional $F$.

Instead of Eq.~(\ref{hp}), in LSPT one begins with the regularized Hamiltonian $H_\Lambda$.  The kink Hamiltonian is defined to be 
\begin{equation}
H\p_\Lambda \equiv \mathcal{D}_f^\dagger H_\Lambda\df.
\label{eq:hpdef}
\end{equation}
Now there is no need to arbitrarily define a regulator $\Lambda\p$ for $H\p_\Lambda$, because as defined $H\p_\Lambda$ is already regularized.  As $H\p_\Lambda$ and $H_\Lambda$ are unitarily equivalent, they have the same spectrum and so one may work with whichever is more convenient at a given step in a problem.  This is in contrast with the $H_\Lambda$ and $H\p_{\Lambda\p}$ that appear in the conventional approaches to quantum solitons, which have different spectra.  As $\Lambda$ is now the only regulator, for simplicity we will leave it implicit from here on.


Note that we do not transform the field from $\phi(x)$ into $\g(x)$.  The field operator $\phi(x)$ does not depend on the sector of the Hilbert space on which it acts.  We always use the same operator.  Furthermore $H$ is always the Hamiltonian which generates time evolution and whose eigenvalues are the energy, no matter which sector of the Hilbert space is considered.

\subsection{States}

Let us consider a vacuum sector state $|\psi\rangle$, which consists of a linear combination of $n$-meson Fock space states up to some finite value of $n$.  Then the expectation value $\langle\psi|\phi(x)|\psi\rangle$ will be of order $O(\sqrt{\hbar})$ as a result of the $\hbar$ in the canonical commutation relations.  Intuitively, the additional $\phi(x)$ leads to half of a commutator, yielding the $\sqrt{\hbar}$.  

What about the state $\df|\psi\rangle$?  The expectation value in this state is
\beq
\langle \psi|\df^\dag \phi(x)\df|\psi\rangle=\langle \psi| \phi(x)+f(x)|\psi\rangle=\langle \psi|\phi(x)|\psi\rangle+f(x)=f(x)+O(\sqrt{\hbar}).
\eeq
In other words, this is a kink sector state.  Inverting the argument, one writes a one-kink sector state as $\df|\psi\rangle$ where $|\psi\rangle$ is in the ordinary Fock space.  The time-evolution of such a state is quite simple.  At time $t$, $\df|\psi\rangle$ becomes
\beq
e^{-iHt}\df|\psi\rangle=\df e^{-iH\p t}|\psi\rangle.
\eeq
In other words, to find the evolution of $\df|\psi\rangle$ it suffices to evolve the ket $|\psi\rangle$ by the action of the kink Hamiltonian $H\p$.

This is good news, but it leads to equations that are somewhat cluttered by the $\df$ at the beginning of every state.  While it is certainly not necessary, to remove the clutter, we change our notation to remove these $\df$ prefactors from the states.  More formally, one can think of the ket vectors as coordinates on our Hilbert space of states.  Every dictionary that identifies the kets with the states is called a frame.  The defining frame is the usual one, in which $H$ time-evolves the states.  We define the kink frame to be the identification of the ket $|\psi\rangle$ with the one-kink sector state that was identified with $\df|\psi\rangle$ in the defining frame.  In other words, we simply remove the $\df$ from the ket without changing the state to which the ket refers.  

One needs to remember, however, that this $|\psi\rangle$ ket is evolved by the kink Hamiltonian $H\p$.  In general relativity, this is the familiar statement that under a passive coordinate transformation one needs to also transform the functions.  Here in quantum theory the statement is that a passive renaming of the kets which are Hilbert space coordinates requires a corresponding renaming of the operators acting on that Hilbert space.

In summary, we will adapt the shorthand notation $|\psi\rangle$ to refer to the state normally called $\df|\psi\rangle$, which will simplify equations below but the price of this simplification is that time evolution is generated by $H\p$.  Again the notation $\phi(x)$ is always used to refer to the same operator, no matter which sector we consider and no matter which frame we choose to define our states.  If desired, one could define a $\phi\p=\phi+f$ analogously to $H\p$, and it would allow us to recover the right expectation value of the field and form factors, but it would not be amenable to perturbative computations because the moments of $\phi\p$ are large even in the vacuum sector.  



The next step is to expand the kink Hamiltonian in powers of the fields~\cite{Evslin_2019}. More specifically, we decompose the kink Hamiltonian into a sum of terms $H'_j$ which, when plane-wave normal ordered, each contain $j$ factors of the operators $\phi$ and $\pi$,
\begin{equation}
\begin{split}
H'_0 &= Q_0, 
\qquad 
H'_1 = 0,
\qquad
H'_2 = \frac{1}{2}\int dx
\Big[:\pi^2(x): + :(\partial_x\phi(x))^2: + :V^{(2)}(\sqrt{\lambda}f(x))\phi^2(x): \Big],\\
H'_{n>2} &= \lambda^{(n-2)/2}\int dx\frac{V^{(n)}(\sqrt{\lambda}f(x))}{n!}:\phi^n(x): 
\end{split}
\end{equation}
where $Q_0$ denotes the classical kink mass.

Just as the free part of the defining Hamiltonian is diagonalized by expanding the fields in plane waves, the quadratic kink Hamiltonian $H'_2$ is diagonalized by expanding the fields in the normal modes of small fluctuations about the kink. These normal modes $\g$ are the constant frequency solutions of the classical equations derived from $H'_2$, and satisfy the Sturm-Liouville eigenvalue problem
\begin{equation}
\left(-\partial_x^2 + V^{(2)}(\sqrt{\lambda}f(x))\right)\mathfrak{g}(x)=\omega^2\mathfrak{g}(x),
\label{eq:SL}
\end{equation}
with a translational zero mode $\mathfrak{g}_B$ at $\omega_B=0$, continuum modes $\mathfrak{g}_k$ with $\omega_k=\sqrt{m^2+k^2}$, and possibly discrete bound (shape) modes $\mathfrak{g}_S$ with $0<\omega_S<m$ which we take to be real. We fix the normalization
\begin{equation}
\int dx \mathfrak{g}^*_{k_1}(x)\mathfrak{g}_{k_2}(x)=2\pi\delta(k_1-k_2),\qquad
\int dx \mathfrak{g}_{S_1}(x)\mathfrak{g}_{S_2}(x)=\delta_{S_1S_2},\qquad
\int dx \mathfrak{g}^2_B(x)=1.
\label{eq:normal}
\end{equation}
We fix the sign of the zero mode and also the decomposition into $\g_{\pm k}$ by the conventions
\begin{equation}
\mathfrak{g}_B(x)=-\frac{f'(x)}{\sqrt{Q_0}}, \qquad \mathfrak{g}^{*}_{k}=\mathfrak{g}_{-k}.
\label{eq:zeromode}
\end{equation}

\subsection{Mode expansion and kink sector Fock space}
\label{subsec:modeexp}
We will now expand the field and its conjugate momentum in the basis of normal modes around the classical kink
\begin{align}\label{phipiNMD}
\phi(x)
&= \phi_{0}\mathfrak{g}_{B}(x)
  + \sum
    \left(
      B_{S}^{\ddagger} + \frac{B_{S}}{2\omega_{S}}
    \right) \mathfrak{g}_{S}(x)
  + \int\frac{dk}{2\pi}
    \left(
      B_{k}^{\ddagger} + \frac{B_{-k}}{2\omega_{k}}
    \right) \mathfrak{g}_{k}(x),
\\
\pi(x)
&= \pi_{0} \mathfrak{g}_{B}(x)
  + i\sum
    \left(
      \omega_{S} B_{S}^{\ddagger} - \frac{B_{S}}{2}
    \right) \mathfrak{g}_{S}(x)
  + i\int\frac{dk}{2\pi}
    \left(
      \omega_{k} B_{k}^{\ddagger} - \frac{B_{-k}}{2}
    \right) \mathfrak{g}_{k}(x).
\end{align}
Here we have adopted the shorthard notation
\beq
B^\ddag_k=\frac{B^\dag_k}{2\ok{}}\hsp B^\ddag_S=\frac{B^\dag_S}{2\os}
\eeq
which will remove some square root factors from the states computed below.

With the normalization conventions of Eq.~(\ref{eq:normal}), the canonical commutation relations obeyed by $\phi(x)$ and $\pi(x)$ imply that the coefficients satisfy the algebra
\begin{equation}\label{BComm}
[B_S,B_{S'}^\ddagger]=\delta_{SS'},\qquad
[B_k,B_{k'}^\ddagger]=2\pi\delta(k-k') \hsp [\phi_0,\pi_0]=i
\end{equation}
with all other commutators vanishing. 
We identify $B_S$ and $B^\ddag_S$ as annihilation and creation operators for the discrete shape modes and $B_k,B_k^\dagger$ as those the continuum, while $\phi_0$ and $\pi_0$ are, at the linear level, proportional to the position and momentum of the kink center of mass.

In the normal-mode basis, the quadratic kink Hamiltonian takes the diagonal form
\begin{equation}
H'_2
=Q_1+\frac{\pi_0^{2}}{2}
+\sum_S \omega_SB_S^\ddagger B_S
+\int\frac{dk}{2\pi}\omega_kB_k^\ddagger B_k ,
\label{eq:H2diag}
\end{equation}
where $Q_1$ is the one-loop correction to the kink mass and $(\phi_0,\pi_0)$ are the canonical zero-mode variables. Thus $H'_2$ describes the center-of-mass motion of the kink as a free particle together with a collection of harmonic oscillators, one for each shape mode $S$ and each continuum mode $k$.

We define the kink sector ground state $|0\rangle_0$ to be the ground state of $H\p_2$, which is
\begin{equation}
B_S|0\rangle_0=0,\qquad B_k|0\rangle_0=0\hsp \pi_0\vac_0=0
\label{eq:kinkvac}
\end{equation}
and construct multiparticle states by acting with creation operators, for example
\begin{equation}
|k_1\cdots k_n\rangle_0 \equiv B_{k_1}^\ddagger\cdots B_{k_n}^\ddagger|0\rangle_0
\label{eq:nmeson}
\end{equation}
where each $k_i$ runs over real values $k$ but may also be a shape mode index $S$.  For example, below we will be interested in the twice-excited shape mode
\beq
|SS\rangle_0=B^\ddag_SB^\ddag_S\vac_0.
\eeq

The zero subscript indicates that these are eigenstates of $H\p_2$, not the full kink Hamiltonian $H\p$.  We denote eigenstates $|\psi\rangle$ of the full Hamiltonian by removing the subscript, but demanding that the leading order term is that above
\beq
|\psi\rangle=\sum_{i=0}^\infty \lambda^{i/2} |\psi\rangle_i.
\eeq
Fixing the leading order part of the state $|\psi\rangle_0$, together with the fact that $|\psi\rangle$ is an eigenstate of the kink Hamiltonian $H\p$ and momentum operator $P\p=\df^\dag P\df$ is not enough to fix a state entirely.  The ambiguity will appear below in the prescriptions for evaluating various poles.

A general kink sector ket, such as $|\psi\rangle_i$, may be expanded in this basis as
\begin{equation}
|\psi\rangle_i
=\sum_{m,n}|\psi\rangle^{mn}_i\hsp |\psi\rangle^{mn}_i=\phi_0^{m}
\int\frac{dk_1\cdots dk_n}{(2\pi)^n}
\gamma^{mn}_i(k_1,\ldots,k_n)|k_1\cdots k_n\rangle_0 .
\label{eq:stateexp}
\end{equation}
Here $|\psi\rangle_i^{mn}$ is the order $\lambda^{i/2}$ component of the state $|\psi\rangle$ that has $m$ zero modes and $n$ mesons or shape modes.  Again, $k$ run over real momenta and also discrete shape modes, and it is understood that the above integration includes a sum over these discrete modes.  The coefficients $\gamma_i^{mn}$ are determined order by order by solving the Schr\"odinger equation for $H\p$, yielding the usual old-fashioned perturbation theory structure with energy denominators from intermediate states \cite{evslin2024reflectioncoefficientreflectionlesskink}.

Turning to interactions, we denote normal ordering with respect to the kink normal modes by $::_b$. This is the normal ordering scheme in which all $B^\ddag$ and $\phi_0$ operators are placed on the left.  Wick's theorem then relates plane-wave normal ordering to normal-mode normal ordering \cite{Evslin_2021normalmode}. In the present work, we will only need the cubic interaction at leading order, which is the first term on the right hand side in
\begin{equation}
H'_3
=\lambda^{1/2}\int dx\frac{V^{(3)}(\sqrt{\lambda}f(x))}{3!}:\phi^3(x):_b + \rm{tadpole}.
\label{H3}
\end{equation}
In later sections, we will specialize to the two shape mode subspace and resum the corresponding bubble diagrams to extract the decay width of the twice-excited shape mode state.

\section{Leading reflection amplitude and the twice-excited shape mode pole}
\label{Sec3}

\begin{figure}[htbp]
\centering
\includegraphics[width = .85\textwidth]{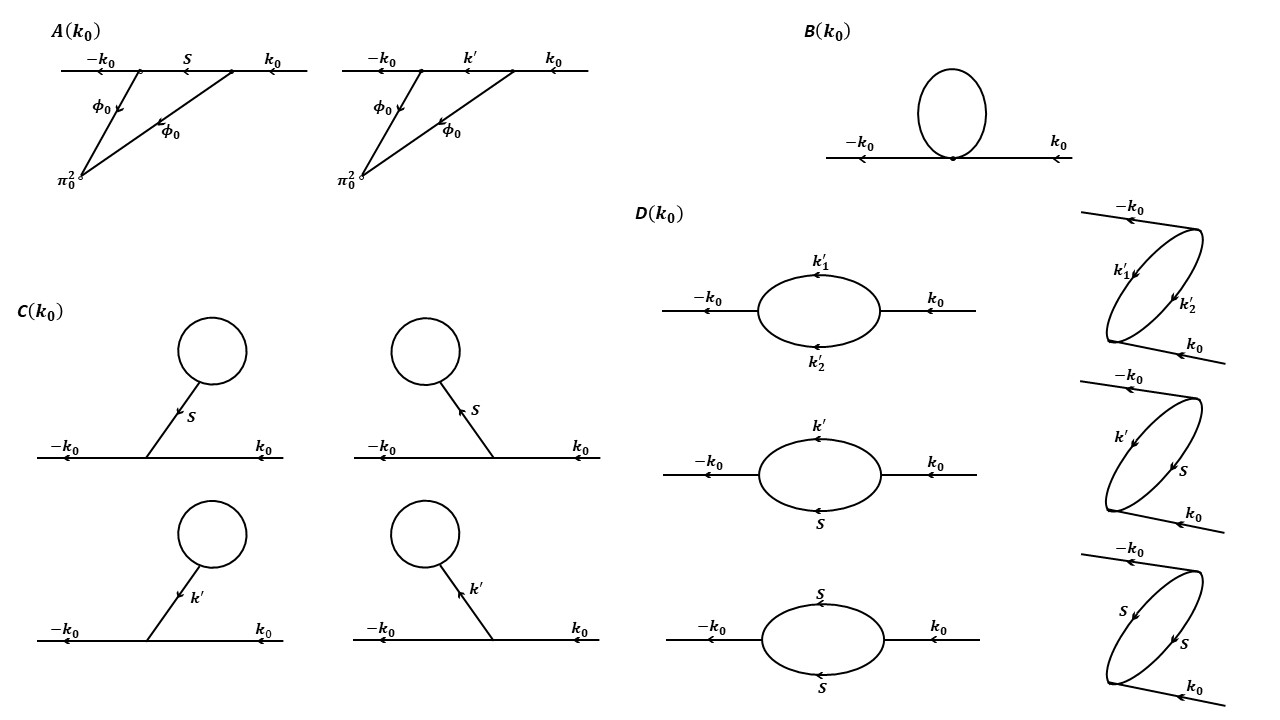}
\caption{The diagrams that compute the components $A$, $B$, $C$ and $D$ of the elastic scattering amplitude $R$  \cite{Bilguun2025elastickinkmesonscatteringphi4}.  Time runs from right to left.  The $\pm k_0$ represent the incoming and outgoing meson momenta, $S$ represents a shape mode, $\phi_0$ is the zero mode and $\pi_0^2$ insertions arise from center of mass kinetic term in $H\p_2$.  The tadpole represents the tadpole term in Eq.~(\ref{H3}).}
\label{phfig1}
\end{figure}
In this section, we review the leading nontrivial (one-loop) reflection amplitude for elastic
kink-meson scattering in LSPT, following Refs.~\cite{evslin2023elastickinkmesonscattering, Bilguun2025elastickinkmesonscatteringphi4}.  We refer to these processes as one-loop, following their diagrammatic representations in Fig.~\ref{phfig1}.  While those diagrams are the most convenient for evaluating their contributions, they differ somewhat from ordinary Feynman diagrams that appear the perturbation theory of particles because the kink is not drawn, and the zero modes and shape modes are discrete degrees of freedom.  The relationship between the two kinds of diagrams is summarized in Table~\ref{tabtab}.  The main difference is that the state diagrams shown here treat shape modes and mesons on equal footing, whereas Feynman diagrams treat shape mode excitations as new varieties of kinks.  As a result, state diagrams are less numerous, their number of loops determines the number of powers of $\lambda\hbar$ in a contribution and they provide the simplest path to calculating the amplitudes below.  However, Feynman diagrams are useful for understanding the pole structure arising from the kinematics.

Our goal in this section is to isolate the pole associated with the twice-excited shape mode $(|SS\rangle)$ intermediate
state.  The goal of Sec.~\ref{Sec4} will be to eliminate this pole.

\subsection{The Lippmann-Schwinger Equation}

In Ref.~\cite{evslin2023elastickinkmesonscattering}, it was shown that the Lippmann-Schwinger equation allows one to read the amplitude for elastic kink-meson scattering directly off of the coefficients $\gamma^{mn}_i$ in Eq.~(\ref{eq:stateexp}).  

If the incoming momentum of a meson in the center of mass frame is $k_0$, so that the outgoing momentum of a reflected meson is $-k_0$, then one may expand the state $|k_0\rangle$ corresponding to one kink and one meson as described above.  Consider the coefficient at order $j$ with a single meson and no zero modes
\beq
\gamma_j^{01}(k\p)=F_j(k_0,k\p)+\frac{R_j(k\p)}{k_0+k\p} \label{rdef}
\eeq
where the functions $F_j(k_0,k\p)$ and $R_j(k\p)$ are both continuous at $k=-k\p$.  There are many ways to define the pole, corresponding to the degenerate eigenstates of the kink momentum and energy, as will be discussed below.  It turns out that $-iR_j(k_0)$ is the order $O(\lambda^{j/2})$ contribution to the matrix element ($i$ times the amplitude) for reflection.  Therefore finding the reflection amplitude is reduced to calculating the states using the perturbation theory described in Sec.~\ref{lsptsez}.

\subsection{Amplitude decomposition at ${\cal O}(\lambda)$}
\label{subsec:ABCD}

The coefficient $\gamma_i^{mn}(k_1\cdots k_n)$ corresponds to the component of the $i$th order correction to an eigenstate of the full interaction kink Hamiltonian $H\p$, projected onto the Fock space corresponding to eigenstates of the free kink Hamiltonian $H\p_2$ with one kink, $m$ zero modes and $n$ mesons or shape modes with momenta or discrete shape mode labels $k_i$.  These contributions may be computed diagrammatically \cite{evslin2023elastickinkmesonscattering} using the state diagrams shown below.  For example, if one is interested in a one-meson state $|k_0\rangle$, then there is always one external line on the right labeled by the momentum $k_0$.  There are $m$ zero mode external lines on the left, which are labeled with a $\phi_0$, and there are $n$ shape mode or meson external lines on the left, which are labeled with the name of the shape mode or momentum.  For example, in the case of $\gamma_i^{01}(k)$, there is only one outgoing line on the left labeled by the momentum $k$. The vertices are the interactions in the Hamiltonian which, in perturbation theory, are used to compute the state.  Proceeding from right to left, one goes to higher orders $i$ in perturbation theory.

The one-loop reflection amplitude can be organized into four contributions,
shown diagrammatically in Fig.~1,
\begin{equation}
R(k_0)=\lambda R_2(k_0)=\lambda\Big(A(k_0)+B(k_0)+C(k_0)+D(k_0)\Big).
\label{eq:RABCD}
\end{equation}
Here we have drawn additional diagrams to specify whether a given internal line is a shape mode or a meson, but the power of these diagrams is that this is not necessary, one needs simply to sum over all meson and shape modes for each internal line.  This is an advantage over Feynman diagrams for the scattering process here, as the excitation of a shape mode would correspond to a line with an excited kink, and so appear quite differently from a virtual meson, despite the fact that their contributions to the states are naturally combined into sums over all nonzero normal modes, as will be evident momentarily.

For concreteness, consider a model with a single shape mode, although the generalization is obvious.  We quote the explicit expressions from Refs.~\cite{evslin2023elastickinkmesonscattering, Bilguun2025elastickinkmesonscatteringphi4},
\bea \label{scat}
A(k_0) &=& \frac{1}{4 \lambda Q_0 k_0}\left[\left(\frac{\omega_{k_0}^2 + \omega_{S}^2}{\omega_{S}}\right) \Delta_{-k_0S} \Delta_{-k_0 S} + \pin{k^\prime}\left(\frac{\omega_{k_0}^2 + \omega_{k^\prime}^2}{\omega_{k^\prime}}\right)\Delta_{-k_0-k^\prime} \Delta_{-k_0 k^\prime}\right], \nonumber \\
B(k_0) &=& \frac{V_{I -k_0 -k_0}}{4 k_0}, \nonumber \\
C(k_0) &=& -\frac{1}{4 k_0}\left[ \frac{V_{IS} V_{S -k_0 -k_0}}{\omega_{S}^2} + \pin{k^\prime} \frac{V_{I k^\prime}V_{-k^\prime -k_0 -k_0}}{{\omega_{k^\prime}^2}}\right], \nonumber \\
D(k_0) &=& \frac{1}{8k_0}\int\frac{dk'_1dk'_2}{(2\pi)^2}\frac{(\omega_{k_1\p} + \omega_{k_2\p}) V_{-k_0 k_1\p k_2\p} V_{-k_0-k_1\p -k_2\p}}{\omega_{k_1\p} \omega_{k_2\p} \left(\omega_{k_0}^2 - \left(\omega_{k_1\p} + \omega_{k_2\p}\right)^2 + i \epsilon\right)} \\ 
&&+ \frac{1}{8k_0}\pin{k^\prime}\frac{(\omega_{k'} + \omega_{S}) V_{-k_0 k'S} V_{-k_0-k'S}}{\omega_{k\p} \omega_{S} \left(\omega_{k_0}^2 - \left(\omega_{k\p} + \omega_{S}\right)^2 + i \epsilon\right)} +\frac{1}{4k_0}\frac{V_{-k_0 S S}V_{-k_0 S S}}{\omega_S \left(  \omega^2_{k_0}- 4 \omega_{S}^2 + i\epsilon\right)}. \nonumber
\eea
If we again use the generalized notation in which $k_i$ may run over the shape modes, then the matrix meson momentum operator has been written in the normal mode basis as
\beq
\Delta_{k_1k_2}=\int dx \g_{k_1}(x)\partial_x\g_{k_2}(x) 
\eeq
and the three-point vertex factor is defined to be
\beq
V_{k_1k_2k_3}=\int dx \V{3}  \g_{k_1}(x)\g_{k_2}(x)\g_{k_3}(x).
\eeq
In this condensed notation, the expressions for $A,\ B$ and $D$ above would be simplified considerably, however our pole involves two shape modes and so would be obscured by this simplification.

A few comments will be useful later. First, Eq.~\eqref{scat} is written for a single
discrete shape mode; in general, a given kink background may support multiple bound modes, in which
case $S$ should be summed. If there are no shape modes, the summands with $S$ indices are not present.

Second, $A$, $B$, and $C$ contribute a smooth, nonresonant background in the kinematic region of
interest. Roughly, $A$ is associated with recoil/zero-mode insertions, $B$ is a local contact term,
and $C$ is a one-loop correction built from mode overlaps and interaction vertices. The term $D$ is
distinguished by the presence of explicit two-particle intermediate states and the corresponding
energy denominators, and it is therefore responsible for threshold behavior and potential resonant
enhancement.  It is the subject of the present work.

\subsection{Channel decomposition of $D$ and the $|SS\rangle$ pole}
\label{subsec:SSpole}

The $D(k_0)$ term contains three distinct contributions
\begin{equation}
D(k_0)=D_{kk}(k_0)+D_{Sk}(k_0)+D_{SS}(k_0),
\label{eq:Ddecompose}
\end{equation}
corresponding to three intermediate states $|k\p_1k\p_2\rangle$, $|Sk\p\rangle$ and $|SS\rangle$.  In the first term, the intermediate state is a ground state kink together with two virtual mesons.  In the second it is a kink with a once-excited shape mode, together with virtual meson.  We will be interested in the third, in which the intermediate state is a kink with a twice-excited shape mode.  Recall that while the once-excited shape mode $|S\rangle$ is stable because $\omega_S<m$, the twice-excited shape mode decays into a ground state kink and a meson, because $2\omega_S>m$.


The first two pieces control the onset of multi-particle channels:
$D_{kk}$ develops the standard two-particle threshold when $\omega_{k_0}\simeq 2m$,
and $D_{Sk}$ turns on when $\omega_{k_0}\simeq \omega_S+m$.
By contrast, the $|SS\rangle$ channel contains the near-resonant denominator
\begin{equation}
\omega_{k_0}^2-4\omega_S^2+i\epsilon,
\label{eq:SSdenom}
\end{equation}
so that as $\omega_{k_0}\to 2\omega_S$ the one-loop expression develops a pole.
Physically this indicates that the intermediate two shape mode configuration becomes nearly on-shell,
and that the scattering amplitude is sensitive to the dynamics of the twice-excited kink state $|SS\rangle$.


\subsection{Feynman Diagrams}

So far, we have used diagrams to compute the Hamiltonian eigenstates in terms of coefficients in a basis of states given by eigenstates of the free Hamiltonian $H\p_2$.  However the Lippmann-Schwinger equation (\ref{rdef}) connects these coefficients with scattering amplitudes.  These scattering amplitudes can be computed using Feynman diagrams.  Therefore, there must be some correspondence between the diagrams above and traditional Feynman diagrams.  While neither set of diagrams is necessary to interpret our results, the conventional bubble diagram interpretation applies to Feynman diagrams, and so we will draw them in addition to our state diagrams.  The conventions used in these diagrams are summarized in Table.~\ref{tabtab}.

\begin{table}[]
\centering
\includegraphics[width=.7\textwidth]{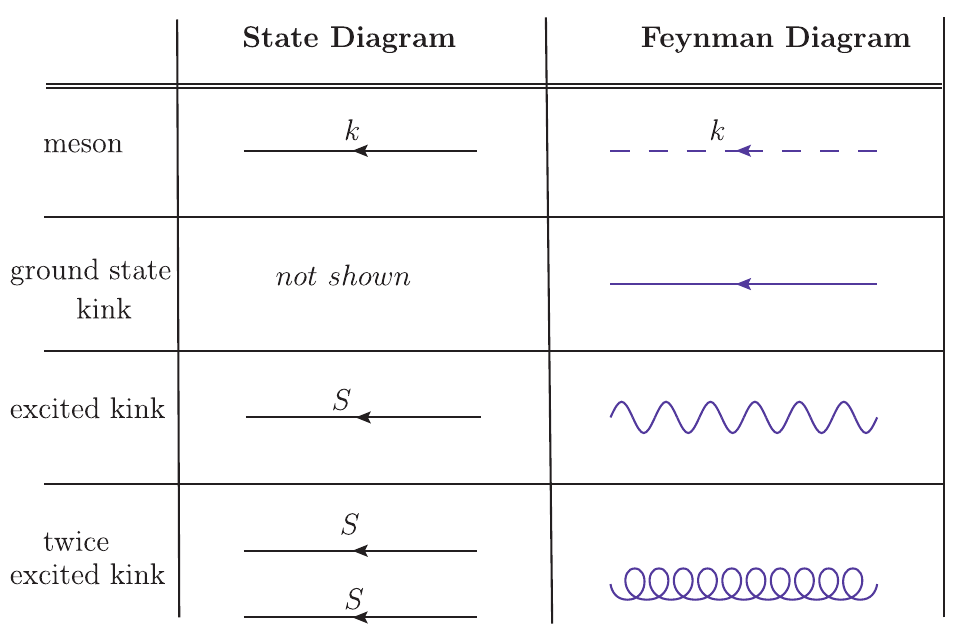}
\caption{Each term in a state can be described by two kinds of diagrams, which we will call state diagrams and Feynman diagrams.  The state diagrams do not explicitly show the kink, but they show shape modes and mesons on equal footing.}
\label{tabtab}
\end{table}



For example, consider the three channels that we have just discussed, corresponding to the contributions $D(k_0)$ in Fig.~\ref{phfig1}.  These all correspond to two diagrams in that figure, although each is drawn three times with different labels to make this clear.  

\begin{figure}[htbp]
\centering
\includegraphics[width=.3\textwidth]{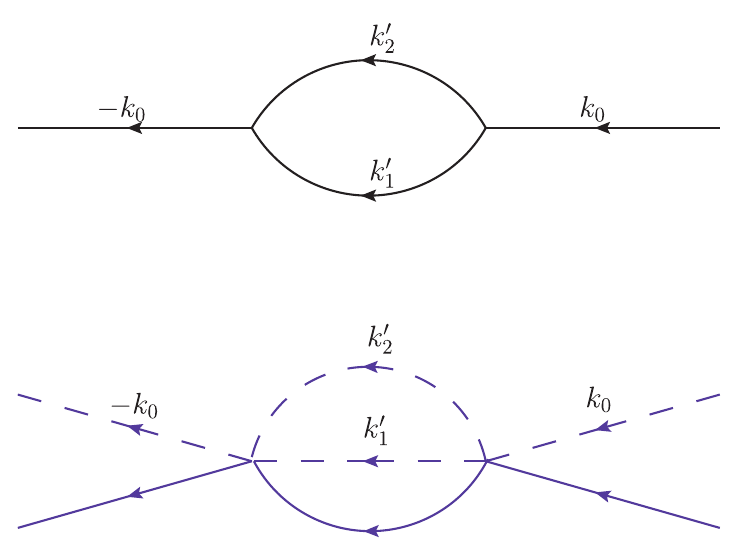}
\includegraphics[width=.3\textwidth]{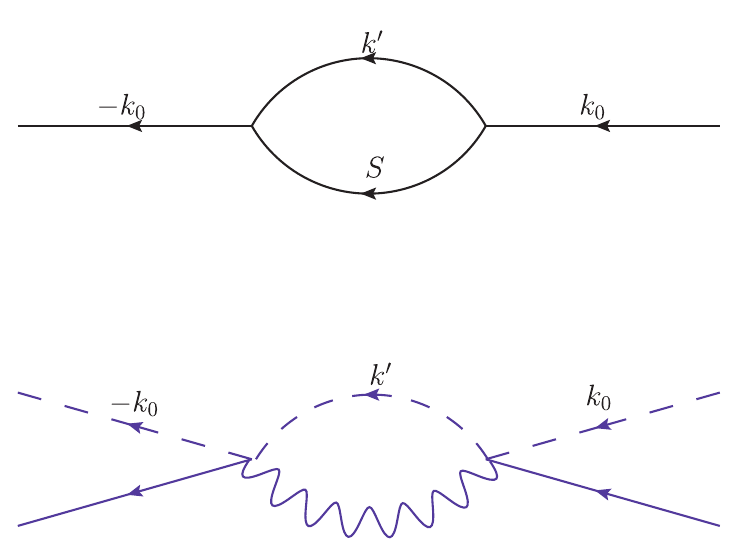}
\includegraphics[width=.3\textwidth]{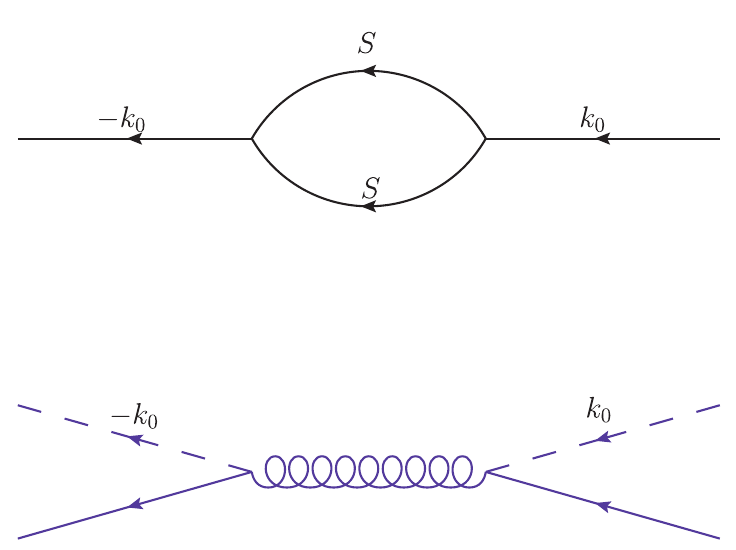}
\caption{The three channels of contributions to $D$ are shown.  The intermediate states have an unexcited kink and two mesons (left), a shape mode excited kink and one meson (center) and a kink with an unstable twice-excited shape mode (right).  The state diagrams are on top, and the corresponding Feynman diagrams on the bottom.  The kink is not shown in the state diagrams.  In the Feynman diagrams, the mesons are drawn as dashed lines while the kink curves are solid and no momentum is given for them, as we work in the center of mass frame.  The kink curve is drawn with more pronounced wiggles when more shape modes are excited.  Note that all three channels correspond to the same state diagram, but the Feynman diagrams have different numbers of loops.   Time always flows to the left.}
\label{fdfig}
\end{figure}

Reading time from right to left, they each correspond to the scattering of a ground state kink and a meson, with three different intermediate states: a doubly excited kink, a singly excited kink with a virtual meson, and finally a ground 
state kink with two virtual mesons.  

In Fig.~\ref{fdfig} these same three identical diagrams are shown above the corresponding Feynman diagrams.  Note that while the corresponding state diagrams each have one loop, reflecting the fact that they arise at the same order in our $\lambda\hbar$ expansion, the corresponding Feynman diagrams have different numbers of loops.  The main difference between the state diagrams and the Feynman diagrams is that each zero mode and shape mode excitation is shown as a curve in the state diagram, while the kink is not shown at all.  On the other hand, in a Feynman diagram, one does not draw a curve corresponding to a shape mode, but rather the kink line is drawn differently depending upon which modes are excited.

As a result, the state diagrams have the advantages that they are simpler, there are less of them, and the states can be read from them directly.  This is because the shape modes and mesons enter the states identically, but Feynman diagrams treat them differently.

\section{The Schr\"odinger picture calculation and the bubbles}
\label{Sec4}

In Sec.~\ref{Sec3} we isolated the two shape mode contribution $D_{SS}(k_0)$ and found the near-resonant enhancement when the incoming meson energy approaches twice the shape mode energy, $\omega_{k_0} \to 2\omega_S$.
In this section we rederive, in the Schr\"odinger picture (old-fashioned perturbation theory in the kink sector), how virtual transitions of the intermediate $|SS\rangle$ configuration into continuum states and back shift the pole off the real axis.
Equivalently, the doubly-excited kink propagator acquires a self-energy with an imaginary part, producing a finite decay width.

We will ignore $H\p_0$, as it is a $c$-number and so does not affect the eigenstates.  We keep only the free $H\p_2$ and the channel in the cubic interaction $H_3\p$ that mixes the two shape mode state into the one-meson continuum and back.
All other channels are analytic at $\omega_{k_0} = 2\omega_S$ and may be treated as a smooth background across the resonance.  It should be understood that all equalities are up to such contributions, so that we obtain only contributions from the leading order pole.  The neglected contributions lead to subleading corrections to the decay width, to which we would like to return in future work.

\subsection{Basis states and the role of projections}
\label{subsec:Sec4_basis}

Recall that the amplitude at each order $i$ is given by the decomposition of the state in Eq.~(\ref{rdef}).  In Sec.~\ref{Sec3} we have already reviewed the leading contribution $R_2$.  What about $R_4$?  This requires us to know the state $|k_0\rangle_4$.  Most terms in $|k_0\rangle_4$ will, at weak coupling, vanish faster than the $O(\lambda)$ contributions that we have already discussed.  The only exception is for values of $k_0$ near the pole, where the contribution to the amplitude may be enhanced by a factor of order $O(1/\lambda)$ or more.  It is in this regime that the corrections discussed below are not subdominant to those reviewed in the previous section. 



We recall that $H_2'$ is diagonal in the kink normal-mode basis
\begin{equation}
H_2'|k\rangle_0 = (Q_1+\omega_k)|k\rangle_0,
\qquad
H_2'|SS\rangle_0 = (Q_1+2\omega_S)|SS\rangle_0,
\qquad
\omega_k = \sqrt{k^2+m^2}.
\label{eq:Sec4_H2_eigs}
\end{equation}
Eq.~(\ref{rdef}) implies that we are only interested in a single coefficient in the state $|k_0\rangle_i$ at each order $i$.  

\subsection{Perturbative expansion of the eigenvalue equation}
\label{subsec:Sec4_expansion}

We construct the Hamiltonian eigenstate $|k_0\rangle$ describing one ground state kink and one meson by solving the
Schr\"odinger equation in the kink frame

\begin{equation}\label{SchEq}
(H_2' + H_3' + H_4' + \cdots)|k_0\rangle = E|k_0\rangle,
\end{equation}
for an eigenstate whose leading term is
$|k_0\rangle_0$. 
We recall that our eigenstate $|k_0\rangle$ is expanded in half-integer powers
\begin{equation}
|k_0\rangle
= |k_0\rangle_0
+ \lambda^{1/2}|k_0\rangle_1
+ \lambda|k_0\rangle_2
+ \lambda^{3/2}|k_0\rangle_3
+ \lambda^2|k_0\rangle_4
+ O(\lambda^{5/2}).
\label{eq:Sec4_state_expansion}
\end{equation}
We also expand the energy, but only in integer powers of $\lambda$,
\begin{equation}
E = E_0 + \lambda E_1 + \lambda^2 E_2 + O(\lambda^3),
\qquad
E_0 = Q_1+\omega_{k_0}.
\label{eq:Sec4_energy_expansion}
\end{equation}
The absence of an $O(\lambda^{n+1/2})$ energy shift follows from the fact that all half-integer powers of $\lambda$ in the Hamiltonian come with odd powers of $\phi$, which shift the zero mode number plus meson number $m+n$ by an odd amount.  As a result, the coefficients $\gamma_i^{mn}$ are nonvanishing only if $i+m+n$ is odd, meaning that the terms with $m$ zero modes and $n$ mesons are weighted by a even power of $\sl$ only in $m+n$ is odd.  The left hand side of the eigenvalue equation $H|k_0\rangle=E|k_0\rangle$, restricted to terms of order $O(\lambda^{i/2})$ with $m$ zero modes and $n$ mesons, is then also nonvanishing only for $i+m+n$ odd, and so since $|k_0\rangle_j$ on the   side of (\ref{SchEq}) is nonvanishing only for $j+m+n$ odd, the whole term is only nonvanishing if $i-j$ is even, corresponding to an integer power $\lambda^{(i-j)/2}$ in the $E$.


The interaction relevant for the $|SS\rangle$ bubble is the cubic kink Hamiltonian which, dropping the tadpole term on the first line and the $|Sk\rangle$ and $|kk\rangle$ channels on the second, is
\bea
H'_3
&=& \frac{\sqrt{\lambda}}{6}\int dx V^{(3)}\big(\sqrt{\lambda}f(x)\big):\phi^3(x):_b
\label{Hp3a}\\
&=&\frac{\sqrt{\lambda}}{2}\intdk V_{SSk}
:\Big(B_S^{\ddagger}+\frac{B_S}{2\omega_S}\Big)^2
\Big(B_k^{\ddagger}+\frac{B_{-k}}{2\omega_k}\Big):_{b},
\label{Hp3b}
\eea
where $k$ is a continuum mode and we remind the reader that the three-point interaction factor is
\begin{equation}
V_{SSk}= \int dx V^{(3)}\big(\sqrt{\lambda}f(x)\big)\mathfrak{g}^2_S(x)\mathfrak{g}_k(x).
\label{VSSk}
\end{equation}

\begin{figure}[htbp]
\centering
\includegraphics[width=.2\textwidth]{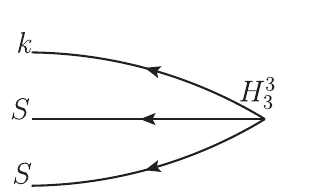}
\includegraphics[width=.25\textwidth]{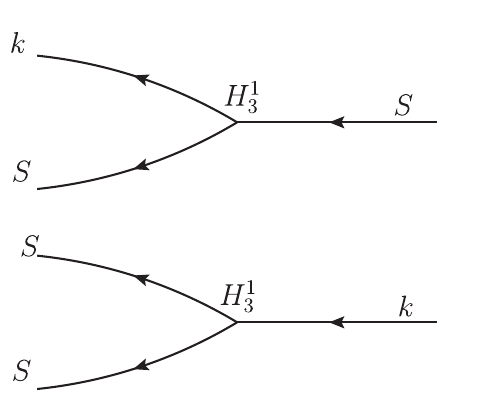}
\includegraphics[width=.25\textwidth]{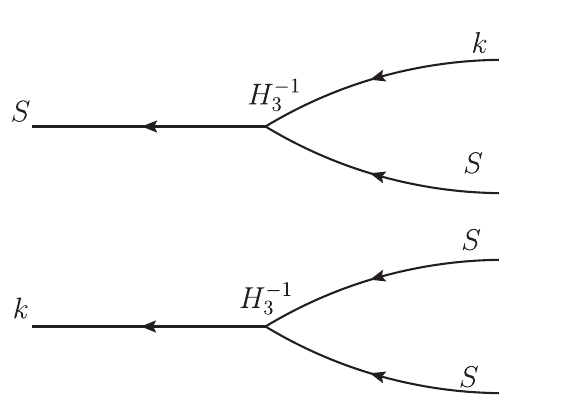}
\includegraphics[width=.2\textwidth]{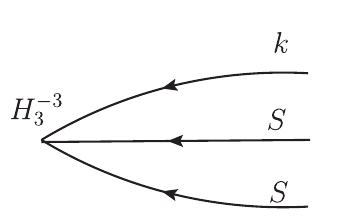}
\caption{The vertices representing the various interactions $H_3^n$ in state diagrams, where $n$ is the meson number.  Note that for each $n$, the diagrams are the same up to the choice of which leg is a shape mode or a meson.  The corresponding Feynman diagrams would depend on how excited the kink is at the beginning.  Recall that we have restricted our attention to interactions with two shape modes and a meson because only these will contribute to the leading order pole.}
\label{h3fig}
\end{figure}

It is convenient to decompose $H'_3$ into summands with fixed meson number,
\begin{equation}
H'_3 = H_3^{3}+H_3^{1}+H_3^{-1}+H_3^{-3},
\end{equation}
with
\begin{align}
H_3^{3}
&=\frac{\sqrt{\lambda}}{2}\intdk V_{SSk}B_S^{\ddagger2}B_k^{\ddagger}, \label{h33}
\\
H_3^{1}
&=\sqrt{\lambda}\intdk V_{SSk}\left(
\frac{B_S^{\ddagger2}B_{-k}}{4\omega_k}
+\frac{B_k^{\ddagger}B_S^{\ddagger}B_S}{2\omega_S}
\right),\label{h31}
\\
H_3^{-1}
&=\sqrt{\lambda}\intdk V_{SSk}\left(
\frac{B_k^{\ddagger}B_S^{2}}{8\omega_S^{2}}
+\frac{B_S^{\ddagger}B_S B_{-k}}{4\omega_S\omega_k}
\right),
\\
H_3^{-3}
&=\frac{\sqrt{\lambda}}{16\omega_S^{2}}\intdk \frac{V_{SSk}}{\omega_k}
B_S^{2} B_{-k}
\end{align}
as depicted in Fig.~\ref{h3fig}.  The superscript indicates the net number of mesons and shape modes created minus those annihilated.  

Substituting the expansions Eqs.~\eqref{eq:Sec4_state_expansion}--\eqref{eq:Sec4_energy_expansion} into Eq.~
\eqref{SchEq} and matching powers of $\lambda^{1/2}$, dropping\footnote{The other interactions do not lead to a maximum number of bubbles and so will lead to subdominant corrections to the width of the resonance.} all contributions arising from interactions other than $H\p_3$, yields the hierarchy
\begin{align}
O(\lambda^0):&\qquad (H_2'-E_0)|k_0\rangle_0 = 0,
\label{eq:Sec4_order0}\\
O(\lambda^{1/2}):&\qquad \sl(H_2'-E_0)|k_0\rangle_1 = -H_3'|k_0\rangle_0,
\label{eq:Sec4_orderhalf}\\
O(\lambda):&\qquad \sl(H_2'-E_0)|k_0\rangle_2 = -H_3'|k_0\rangle_1 + E_1|k_0\rangle_0,
\label{eq:Sec4_order1}\\
O(\lambda^{3/2}):&\qquad \sl(H_2'-E_0)|k_0\rangle_3 = -H_3'|k_0\rangle_2 + E_1|k_0\rangle_1,
\label{eq:Sec4_order3half}\\
O(\lambda^2):&\qquad \sl(H_2'-E_0)|k_0\rangle_4 = -H_3'|k_0\rangle_3 + E_1|k_0\rangle_2 + E_2|k_0\rangle_0.
\label{eq:Sec4_order2}
\end{align}
The energy shifts $E_1$ and $E_2$ are smooth functions of $\omega_{k_0}$ and do not control the
resonant enhancement near $\omega_{k_0}\simeq 2\omega_S$. The Breit-Wigner behavior instead
originates from the explicit two-particle denominators associated with intermediate $|SS\rangle$ states,
which we will extract in the following subsections.

\subsection{Solving the eigenvalue equation for $|k_0\rangle$}
\label{subsec:Sec4_statecorr}

\subsubsection{$O(\lambda^0)$: Fixing $E_0$}
At leading order, we have defined
\beq
|k_0\rangle_0=B^\ddag_k\vac_0 \label{k0def}
\eeq
and our eigenvalue equation is Eq.~(\ref{eq:Sec4_order0})
\beq
H\p_2|k_0\rangle_0=\left(Q_1+\omega_{k_0}\right)|k_0\rangle_0=E_0|k_0\rangle_0
\eeq
which is satisfied when we set
\beq
E_0=Q_1+\omega_{k_0}.
\eeq

\subsubsection{$O(\lambda^{1/2})$: The $|SS\rangle$ admixture and the resonant denominator}

Let us begin with the right hand side of Eq.~\eqref{eq:Sec4_orderhalf}, keeping only the terms in Eqs.~(\ref{h33}) and~(\ref{h31}) 
\bea
-H_3\p|k_0\rangle_0&=&-H_3^3|k_0\rangle_0-H_3^1|k_0\rangle_0\\
&=&-\frac{\sqrt{\lambda}}{2}\intdk V_{SSk}B_S^{\ddagger2}B^\ddag |k_0\rangle_0-{\sqrt{\lambda}}\intdk V_{SSk}\left(
\frac{B_S^{\ddagger2}B_{-k}}{4\omega_k}
\right)|k_0\rangle_0\nonumber\\
&=&-\frac{\sqrt{\lambda}}{2}\intdk V_{SSk} |SSkk_0\rangle_0-\frac{\sqrt{\lambda}}{4
\ok{0}}V_{SS-k_0}|SS\rangle_0.\nonumber
\eea
Other contributions would be finite as $k$ approaches $k_0$ and so will not contribute to our residue $R$.  To avoid clutter, we do not include them.

What about the left hand side?  As reported in Eq.~(\ref{eq:H2diag}) $H\p_2$ has several terms.  The $\pi_0^2/2$ term is a kinetic term for the center of mass of the kink.  It is not invertible.  This issue is explained in Ref.~\cite{Evslin_2020}.  We are interested in a state that is an eigenvector not only of the kink Hamiltonian $H\p$ but also of the kink momentum $P\p$.  This second condition fixes all terms $\gamma_i^{mn}$ with zero mode number $m>0$.   With these fixed, $H\p_2$ becomes invertible up to terms that fix the prescription for evaluating the poles.

On the right hand side, we have only found terms with two shape modes, either zero or two mesons, and no zero modes.  Let us match these to the same terms on the left hand side.  Since none of the terms in $H\p_2$ changes the number of mesons, we can treat sectors with different numbers of mesons separately.

Let us start with the first term on the right hand side.  There are two such contributions on the left hand side
\beq
\sl\left(Q_1
+\sum_S \omega_SB_S^\ddagger B_S
+\int\frac{dk}{2\pi}\omega_kB_k^\ddagger B_k-E_0\right)|k_0\rangle_1^{04}+\sl\left(\frac{\pi_0^{2}}{2}\right)|k_0\rangle_1^{24}.
\eeq
In the first term, there are no zero modes in the operator or in the state $|k_0\rangle_0^{04}$ whereas the second term contains the part of the state which has two zero modes, but which are annihilated by the kink center of mass kinetic term $\pi_0^2/2$.

\begin{figure}[htbp]
\centering
\includegraphics[width=.4\textwidth]{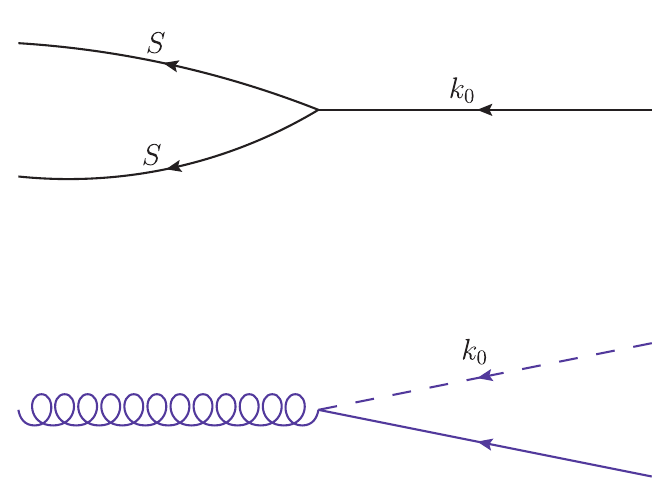}
\includegraphics[width=.4\textwidth]{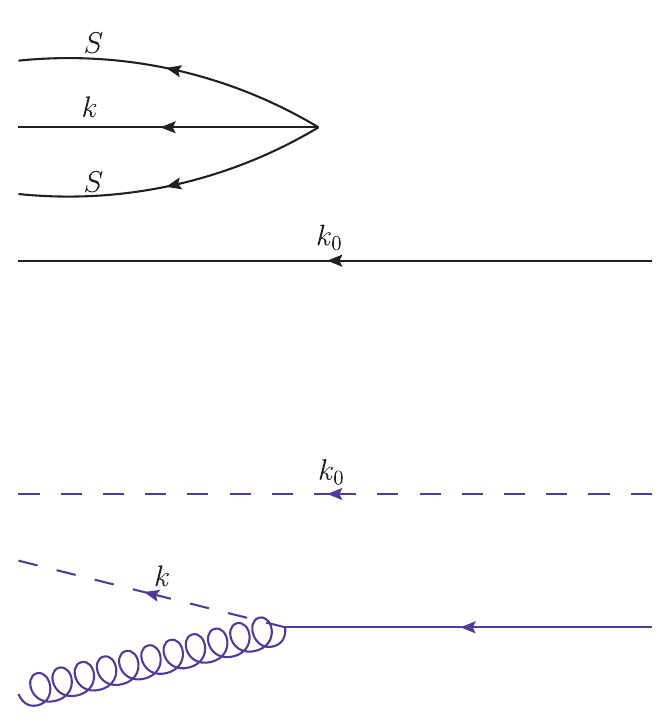}
\caption{The two channels contributing to $|k_0\rangle_1^{02}$ (left) and $|k_0\rangle_1^{04}$ (right) respectively.  The state diagram is shown on top and the Feynman diagram on the bottom. Note that $|k_0\rangle_1^{04}$ arises from virtual shape modes and a virtual meson which are also present in the dressed kink.  
$|k_0\rangle_1^{02}$ arises from a meson that is absorbed by a kink, twice exciting its shape mode.}
\label{k1fig}
\end{figure}

As we know $E_0$ and we know that the right hand side is proportional to $|SSkk_0\rangle_0$, the left hand side can be simplified somewhat to
\beq
\sl\left(2\omega_S-\ok{0}+\int\frac{dk}{2\pi}\omega_kB_k^\ddagger B_k\right)|k_0\rangle_1^{04}+\sl\left(\frac{\pi_0^{2}}{2}\right)|k_0\rangle_1^{24}.
\eeq
The first term already has a suggestive coefficient, suggesting that it may give rise to a pole.  The second does not, and so it will not contribute to $R$.  Moreover, the momentum operator produces a total shift of $m$ and $n$ of 2 units at each order, and so $|k_0\rangle^{24}_1$ vanishes for a state which is annihilated by $P\p$, in other words by a momentum eigenstate in the center of mass frame.  Therefore the second term is equal to zero.  We are left with
\beq
\left(2\omega_S-\ok{0}+\int\frac{dk}{2\pi}\omega_kB_k^\ddagger B_k\right)|k_0\rangle_1^{04}=-\frac{1}{2}\intdk V_{SSk} |SSkk_0\rangle_0
\eeq
which is easily solved to yield
\beq
|k_0\rangle_1^{04}=-\frac{1}{2}\intdk \frac{V_{SSk}}{(2\omega_S+\ok{})} |SSkk_0\rangle_0 \label{k14}
\eeq
or equivalently
\beq\label{gamma104}
\gamma_1^{04}(k)=-\frac{V_{SSk}}{2(2\omega_S+\ok{})}
\eeq
where we remind the reader that we have dropped many terms that will not eventually lead to the desired pole.  Note that the denominator is always positive, as the two virtual shape modes and the virtual meson of momentum $k$ never conserve energy and so are quite off-shell.  The calculation above is summarized diagrammatically in the right panel of Fig.~\ref{k1fig}.

Now let us proceed to the other term on the right hand side, with two shape modes and no mesons, which is shown in the left panel of Fig.~\ref{k1fig}.  Again $|k_0\rangle_1^{22}$ vanishes and proceeding as above we find
\beq
\left(2\omega_S-\ok{0}+\int\frac{dk}{2\pi}\omega_kB_k^\ddagger B_k\right)|k_0\rangle_1^{02}=-\frac{1}{4\ok{0}}V_{SS-k_0}|SS\rangle_0
\eeq
which is solved by
\beq
|k_0\rangle_1^{02}=-\frac{V_{SS-k_0}}{4\ok{0}(2\omega_S-\ok{0})}|SS\rangle_0 \label{k12}
\eeq
or equivalently
\beq\label{gamma102}
\gamma_1^{02}(k)=-\frac{V_{SS-k_0}}{4\ok{0}(2\omega_S-\ok{0})}.
\eeq
This coefficient is the amplitude to find a virtual twice-excited shape mode inside a system consisting of a ground state kink and a meson.  As usual, it has a pole when the total mass of the virtual system is equal to that of the original system.   The position of the pole is that of the pole in $D_{SS}$.

How is the pole to be defined?  Do we need to add $i\epsilon$?  This is a rather subtle point.  The state $|SS\rangle_0$ in fact is not the leading order expansion of any Hamiltonian eigenstate $|SS\rangle$ because it is unstable.  As a result of this instability, $|SS\rangle_0$ cannot be the initial or final state of such an interaction, although it is not evident at this order in perturbation theory, and so the usual causality argument requiring the $i\epsilon$ does not apply. 

\subsubsection{$O(\lambda)$: $|SS\rangle$ goes into the one-meson continuum}
Now let us proceed to order $i=2$, keeping only the terms from Eqs.~(\ref{k14}) and (\ref{k12}), as only these correspond to the production of virtual pairs of shape modes, although the former also contains two virtual mesons.

Let us warm up by trying to reproduce the pole in Subsec.~\ref{subsec:SSpole}.  For this, we have seen that we need to evaluate $\gamma_2^{01}$, and so we need the one-meson part of $|k_0\rangle_2$.

On the right hand side of Eq.~\eqref{eq:Sec4_order1}, there are two terms that have meson number equal to one
\beq
-H_3^{-3}|k_0\rangle_1^{04}-H_3^{-1}|k_0\rangle_1^{02}.
\eeq
In the first term there are two possible contractions.  The meson annihilation operator in the $H_3^{-3}$ could annihilate the meson with momentum $k$ or that with momentum $k_0$.  The first case would create a disconnected diagram, and so it will certainly not shift our pole.  We therefore restrict attention to the second contraction, simply dropping the first.

With this simplification, the first term on the right hand side becomes
\beq
\frac{\sqrt{\lambda}}{16\omega_S^{2}}\frac{V_{SS-k_0}}{\omega_{k_0}}
 \intdk \frac{V_{SSk}}{(2\omega_S+\ok{})} |k\rangle_0
\eeq
while the second is
\beq
\frac{\sqrt{\lambda}V_{SS-k_0}}{16\ok{0}(2\omega_S-\ok{0})}\intdk 
\frac{V_{SSk}}{\omega_S^{2}}
|k\rangle_0. \label{rhs2}
\eeq
They are quite similar, but the second has a pole and the first does not.  This is because the first term described a system that had not only a virtual twice-excited shape mode, but also two virtual mesons, and so can never be on-shell.  Clearly this first term will be subdominant by one power of $(2\omega_S-\ok 0)$ and so at leading order will not affect our residue $R$.  Therefore, we will keep only the second term.  

This leads us to an important simplification which will help us at higher orders.  Even though the diagrams leading to both terms contained a single doubly-excited meson, that in which the double excitation is created when the meson is destroyed, corresponding to $H^1_3$ is the most divergent because it has the least virtual particles, as is that in which the double excitation is destroyed when a meson is created, corresponding to $H^{-1}_3$.  The interactions $H^3_3$ and $H^{-3}_3$ lead to lower order poles, and so we will drop them.

What is the left hand side of Eq.~\eqref{eq:Sec4_order1} which is equal to Eq.~(\ref{rhs2})?  Again there are two contributions
\beq
\sl\left(Q_1
+\sum_S \omega_SB_S^\ddagger B_S
+\int\frac{dk}{2\pi}\omega_kB_k^\ddagger B_k-E_0\right)|k_0\rangle_2^{01}+\sl\left(\frac{\pi_0^{2}}{2}\right)|k_0\rangle_2^{21}.
\eeq
As we are interested in terms on the right hand side with a single meson and no shape modes, we may simplify this to
\beq
\sl\left(-\ok{0}
+\int\frac{dk}{2\pi}\omega_kB_k^\ddagger B_k\right)|k_0\rangle_2^{01}+\sl\left(\frac{\pi_0^{2}}{2}\right)|k_0\rangle_2^{21}.
\eeq
One can see that the first term will lead to another pole for $k\sim \pm k_0$, which will see is necessary for this to contribute to leading order residue.  The second term, is proportional to $|k_0\rangle_2^{21}$, which is fixed by translation invariance, not dynamics, and can be shown to have no such pole.  Therefore we will drop the second term.

The eigenvalue equation \eqref{eq:Sec4_order1} then becomes
\beq
\left(-\ok{0}
+\int\frac{dk}{2\pi}\omega_kB_k^\ddagger B_k\right)|k_0\rangle_2^{01}=\frac{V_{SS-k_0}}{16\ok{0}(2\omega_S-\ok{0})}\intdk 
\frac{V_{SSk}}{\omega_S^{2}}
|k\rangle_0
\eeq
which is solved by
\beq
|k_0\rangle_2^{01}=\frac{V_{SS-k_0}}{16\ok{0}\omega_S^2(\ok{0}-2\omega_S)}\intdk 
\frac{V_{SSk}}{\ok{0}-\ok {}+i\epsilon}
|k\rangle_0 \label{k2}
\eeq
corresponding to
\beq
\gamma_2^{01}(k)=\frac{V_{SS-k_0}}{16\ok{0}\omega_S^2(\ok 0-2\omega_S)}
\frac{V_{SSk}}{\ok 0-\ok{}+i\epsilon}. \label{g2eqb}
\eeq
This contribution corresponds to the third panel in Fig.~\ref{fdfig}.  As can be seen in that Feynman diagram, the $\ok{0}-2\os$ denominator corresponds to the usual S-channel pole in kink-meson scattering, not to the self-energy of the twice-excited kink.  Indeed, this Feynman diagram does not have any loops.

Here the $i\epsilon$ follows the usual {\textit{in}} state prescription, which chooses the initial condition
\beq
|t=0\rangle=\pin{k\p} e^{-\sigma^2(k\p-k_0)^2-i(k\p-k_0)x_0}|k\p\rangle\hsp 0<1/k_0\ll \sigma\ll -x_0 
\eeq
corresponding to a wave packet of Hamiltonian eigenstates such that there is no reflection before the scattering.  To see this, note that the $O(\lambda)$ one-meson part of the initial wave packet is
\bea
|t=0\rangle_2^{01}&=&\pin{k\p} e^{-\sigma^2(k\p-k_0)^2-i(k\p-k_0)x_0}|k\p\rangle_2^{01}\label{initl}\\
&=&\intdk\left[\pin{k\p} e^{-\sigma^2(k\p-k_0)^2-i(k\p-k_0)x_0}
\frac{V_{SS-k\p}}{16\okp{}\omega_S^2(\okp{}-2\omega_S)} 
\frac{V_{SSk}}{\okp{}-\ok {}+i\epsilon}
\right]|k\rangle_0\nonumber.
\eea
As $x_0$ is very negative, the $k\p$ integration can be closed in the upper imaginary part of the complex plane.  The wave packet can be taken to be sufficiently monochromatic so that the $\okp{}-2\os$ pole's residue is suppressed by the Gaussian.  However, as $k$ is itself a dummy variable, there will always be values of $k$ in the integrand so that the $\okp{}-\ok{}$ poles need to be considered.  As $k_0$ is positive, the Gaussian will suppress the residue when $k\p$ is negative.  However the $\okp{}-\ok{}$ term yields poles at both $k=\pm k\p$.  When $k$ is negative, this denominator is proportional to $k\p+k+i\epsilon$ so that the pole is in the negative imaginary part of the complex plane, and so is not in our contour.  Therefore this $+i\epsilon$ prescription guarantees that we do not have any left-going mesons in our initial state, as desired.  This is the reason that we choose $+i\epsilon$.  
What about right-going mesons in the initial state?  Obviously they appear at leading order.  But at order $O(\lambda)$, another contribution arises from Eq.~(\ref{initl}), as the corresponding pole is proportional to $k\p-k+i\epsilon$.  Again the pole is in the negative imaginary part of the complex plane, and so there is no contribution at initial times.  If there had been such a contribution, it could anyway be absorbed into the normalization of the initial state.

This concludes our discussion of the initial wave packet, now we return to the construction of the Hamiltonian eigenstate.

Near the $k=-k_0$ pole, we may approximate
\beq
\frac{1}{\ok 0-\ok{} +i\epsilon}=\frac{\ok 0}{k_0}\frac{1}{k_0+k+i\epsilon}
\eeq
and so the residue of the $(k+k_0)$ pole, when $2\omega_S\sim\ok{0}$, is therefore
\beq
R_2(k_0)=\frac{V^2_{SS-k_0}}{16k_0\omega_S^2(\ok{0}-2\omega_S)}=\frac{V^2_{SS-k_0}}{4k_0\omega_S(\ok{0}^2-4\omega^2_S
)}
\eeq
in agreement with the pole term in Eq.~(\ref{scat}), which was taken from Ref.~\cite{evslin2023elastickinkmesonscattering}. 

We will drop the terms which are finite at the $k=\pm k_0$ poles, and so simply write
\beq
|k_0\rangle_2=\int \frac{dk}{2\pi} \gamma_2^{01}(k)|k\rangle_0\hsp \gamma_2^{01}(k)=\left(\frac{k_0R_2(k_0)}{\ok{0}}\right)\frac{1 }{\ok{0}-\ok{}+i\epsilon}. \label{g2eq}
\eeq

Let us consider, for the moment, the pole at $k=-k_0$, corresponding to reflection.  Keeping only the pole term, we find
\beq
\gamma_2^{01}(k)=\frac{R_2(k_0)}{k+k_0+i\epsilon}.
\eeq
As usual, there are also terms which are finite at $k=k_0$, which for brevity are not included in our expressions.


\subsubsection{$O(\lambda^{i+1/2})$: Return to $|SS\rangle$} 

Imagine that at some order $i$ we have found
\beq
|k_0\rangle_{2i}=\intdk \gamma_{2i}^{01}(k)|k\rangle_0\hsp \gamma_{2i}^{01}(k)=\left(\frac{k_0R_{2i}(k_0)}{\ok{0}}\right)\frac{1 }{\ok{0}-\ok{}+i\epsilon}
\eeq
plus terms that are finite at the poles.   Here the $k=-k_0$ pole is defined, as above, by the $k+k_0+i\epsilon$ prescription.  Imagine further that at each lower value of $i$, there is an $i$th order pole at $k=\pm k_0$. The case $i=1$ appears in Eq.~(\ref{g2eq}). 

Now let us proceed to calculate $\gamma_{2i+1}$ using the eigenvalue equation
\beq
\sl(H_2'-E_0)|k_0\rangle_{2i+1} = -H_3'|k_0\rangle_{2i} . \label{receq}
\eeq

\begin{figure}[htbp]
\centering
\includegraphics[width=.7\textwidth]{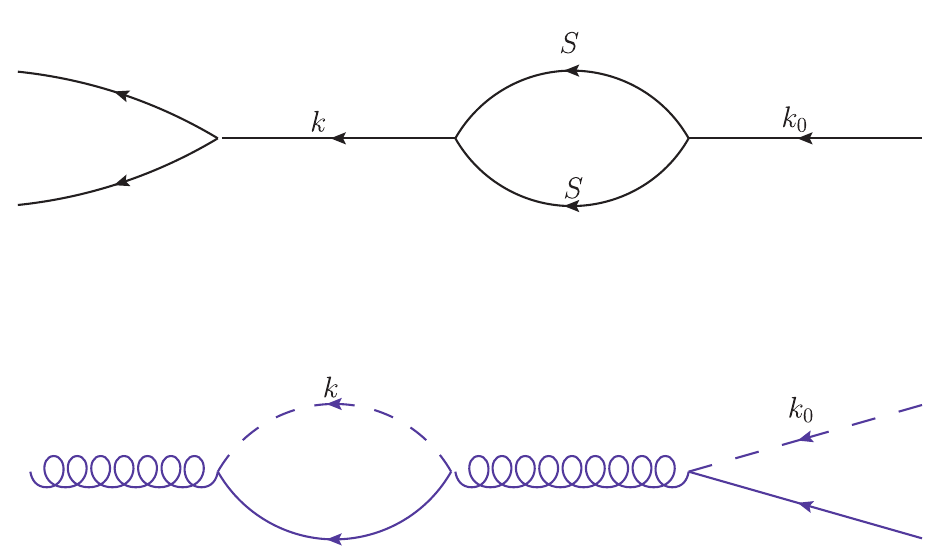}
\caption{The state (top) and Feynman (bottom) diagrams describing the contributions to $|k_0\rangle_3$ in Eq.~(\ref{k32}).  The intermediate kink plus meson configuration corresponds to a loop in the Feynman diagram, which leads to the $\ok{}-\ok{0}$ pole.  However it corresponds to a line in the state diagram, as the kink is not drawn.  The loop in the state diagram leads to the $\ok{0}-2\os$ pole.  More generally, the contribution to $|k_0\rangle_{2i+1}$ would consist of $i$ loops separated by internal lines, repeating the pattern seen here.}
\label{k32fig}
\end{figure}

We have learned above that the highest order pole will arise from the $H_3^1$ term in $H_3\p$.  The leading pole on the right hand side is therefore
\beq
-H_3^1 |k_0\rangle_{2i}=- \intdk \gamma_{2i}^{01}(k) H_3^1 |k_0\rangle_0=-\frac{\sl}{4}\intdk \frac{\gamma_{2i}^{01}(k)V_{SS-k}}{\ok{}}|SS\rangle_0.
\eeq
The eigenvalue equation (\ref{receq}) then yields
\beq
|k_0\rangle_{2i+1}= \intdk \gamma_{2i}^{01}(k) H_3^1 |k_0\rangle_0=\frac{1}{4(\ok{0}-2\os)}\intdk \frac{\gamma_{2i}^{01}(k)V_{SS-k}}{\ok{}}|SS\rangle_0 \label{k32}
\eeq
as is drawn in Fig.~\ref{k32fig}.  

\subsubsection{$O(\lambda^{i+1})$} 

At the next order, we will apply the eigenvalue equation
\beq
\sl(H_2'-E_0)|k_0\rangle_{2i+2} = -H_3'|k_0\rangle_{2i+1} + \sum_{j=1}^{i-1} E_j|k_0\rangle_{2(i+1-j)} . \label{receqb}
\eeq
 By assumption, the $E_j$ terms have lower order poles, and so will not affect the leading pole calculation here, so we will drop them.  Now we have learned that the dominant pole on the right hand side will arise from $H_3^{-1}$, and so the right hand side is
 \beq
 -H_3^{-1}|k_0\rangle_{2i+1}=\frac{\sl}{16\os^2(2\os-\ok{0})}\pin{k\p} \intdk \frac{\gamma_{2i}^{01}(k)V_{SS-k}V_{SSk\p}}{\ok{}}|k\p\rangle_0.
 \eeq

The eigenvalue equation then implies
\beq
|k_0\rangle_{2i+2}=\frac{1}{16\os^2(\ok{0}-2\os)}\intdk\pin{k\p} \frac{\gamma_{2i}^{01}(k)V_{SS-k}V_{SSk\p}}{\ok{}(\ok{0}-\okp{}+i\epsilon)}|k\p\rangle_0 \label{k41}
\eeq
corresponding to
\bea
\gamma_{2i+2}^{01}(k)&=&\frac{V_{SSk}}{16\os^2(\ok{0}-2\os)(\ok{0}-\ok{}+i\epsilon)}\pin{k\p} \frac{V_{SS-k\p}}{\okp{}}\gamma_{2i}^{01}(k\p)\label{passo}\\
&=&\frac{V_{SSk}}{4\os(\ok{0}^2-4\os^2)(\ok{0}-\ok{}+i\epsilon)}\pin{k\p} \frac{V_{SS-k\p}}{\okp{}}\gamma_{2i}^{01}(k\p)\nonumber
\eea
where in the last line we made the approximation $\ok{0}\sim 2\os$, which is valid close to the pole.   

\begin{figure}[htbp]
\centering
\includegraphics[width=.7\textwidth]{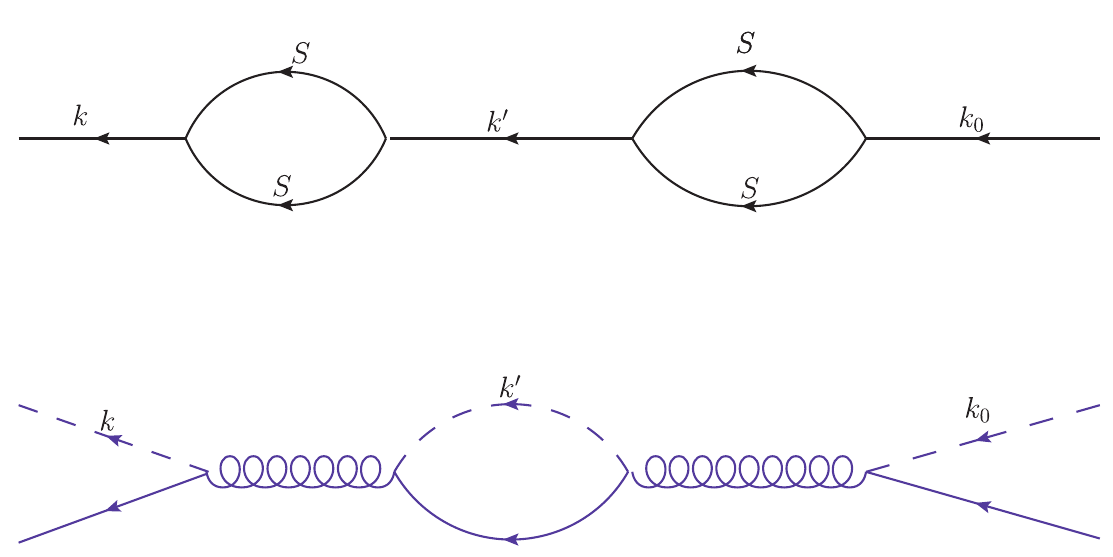}
\caption{The state (top) and Feynman (bottom) diagrams describing the contributions to $|k_0\rangle_4$ in Eq.~(\ref{k41}). }
\label{k41fig}
\end{figure}

Note that
\beq
\gamma_{0}^{01}(k\p)=2\pi\delta(k\p-k_0)
\eeq
as a result of the definition of $|k_0\rangle_0$ in Eq.~(\ref{k0def}), and so in the case $i=0$ this reproduces Eq.~(\ref{g2eqb}).  The case $i=1$ is drawn in Fig.~\ref{k41fig}.

Inserting Eq.~(\ref{passo}) into itself recursively, one finds
\bea
\gamma_{2i+2}^{01}(k)&=&\frac{V_{SSk}}{4\os(\ok{0}^2-4\os^2)(\ok{0}-\ok{}+i\epsilon)}I(k_0)^j\pin{k\p} \frac{V_{SS-k\p}}{\okp{}}\gamma_{2(i-j)}^{01}(k\p)\\
I(k_0)&=&\pin{k}\frac{|V_{SSk}|^2}{4\os\ok{}(\ok{0}^2-4\os^2)(\ok{0}-\ok{}+i\epsilon)}
\nonumber
\eea
for any nonnegative integer $j\leq i$.  Here we have used $V_{SSk}^*=V_{SS-k}$ which results from the reality of the $V^{(3)}$ and $\g_S$ together with the convention $g_k^*=g_{-k}$.

Setting $j=i$ this reduces to
\beq
\gamma_{2i+2}^{01}(k)=\frac{V_{SSk}V_{SS-k_0}}{4\ok{0}\os(\ok{0}^2-4\os^2)(\ok{0}-\ok{}+i\epsilon)}I(k_0)^i.
\eeq

The corresponding total correction to the one-meson Fock space component of $|k_0\rangle$ is the sum of these contributions
\bea
\sum_{i=1}^\infty \lambda^i \gamma_{2i}^{01}(k)=\frac{\lambda V_{SSk}V_{SS-k_0}}{4\ok{0}\os(\ok{0}^2-4\os^2)(\ok{0}-\ok{}+i\epsilon)}\frac{1}{1-\lambda I(k_0)}.
\eea


\subsection{The Decay Rate}

What is $I(k_0)$?  It looks real, except for the $i\epsilon$ in the pole prescription.  Therefore, the imaginary contribution corresponding to reflection arises from the residues at the $k\p=\pm k_0$ poles
\bea
\frac{1}{\ok{0}-\ok{}+i\epsilon}&=&\frac{\ok 0}{k_0}
\left(\frac{1}{k_0+k+i\epsilon}+\frac{1}{k_0-k+i\epsilon}\right)+\rm{finite}\\
&=&-\frac{\ok 0}{k_0} i \pi\left(\delta(k_0+k)+\delta(k_0-k)\right)+\rm{finite}\nonumber
\eea
where the finite part is to be integrated using the Cauchy Principal Value prescription around the $k=\pm k_0$ poles
\beq
\Re[I(k_0)]={\rm{PV}}[I(k_0)].
\eeq

In all, the imaginary part is
\beq
\Im[I(k_0)]=-\frac{2|V_{SSk_0}|^2}{8\os k_0 (\ok{0}^2-4\os^2)}
\eeq
where the factor of two in the numerator came from the sum over the two poles.  The factor of 2 in the denominator came from the $i\pi$ in the Sokhotski–Plemelj theorem divided by the $2\pi$ in the $dk/(2\pi)$.  
The sum of the geometric series is
\beq
\frac{1}{1-\lambda\left({\rm{PV}}[I(k_0)]+i \lambda \Im[I(k_0)]\right)}=\frac{\ok{0}^2-4\os^2}{\ok{0}^2-4\os^2+i\lambda |V_{SSk_0}|^2/(4\os k_0) }=\frac{\ok{0}-2\os}{\ok{0}-2\os+i\Gamma/2}
\eeq
where in the first equality we have dropped the principal value of $I(k_0)$, which does not shift the imaginary part of the pole
.  The last equality matches this complex pole with the Breit-Wigner form, with decay rate
\beq\label{Gamma}
\Gamma=\frac{\lambda |V_{SSk_0}|^2}{8\os^2 k_0}.
\eeq
This is the usual Breit-Wigner pole shift corresponding to a decay of the $|SS\rangle$ state into a left-traveling meson, as it arises from the $k\p=\pm k_0$ poles and $k_0>0$.  
This indeed is the decay rate found in quantum field theory in Ref.~\cite{Evslin_2022}, which agrees precisely with the classical result of Ref.~\cite{mm}.

\subsection{The Residue}

The total correction to the one-meson state is now
\bea
\sum_{i=1}^\infty \lambda^i \gamma_{2i}^{01}(k)=\frac{\lambda V_{SSk}V_{SS-k_0}}{\ok{0}(\ok{0}-\ok{}+i\epsilon)}\left(\frac{k_0}{4k_0\os(\ok{0}^2-4\os^2)+i\lambda |V_{SSk_0}|^2 }\right).
\eea
Matching this to the definition Eq.~(\ref{rdef}) of the residue $R$ at the $k_0+k$ pole, one finds
\beq
R=\frac{\lambda |V_{SSk_0}|^2}{4k_0\os(\ok{0}^2-4\os^2)+i\lambda |V_{SSk_0}|^2 }.
\eeq

The imaginary part of the residue is then
\beq
\Im[R]=-\frac{\lambda^2  |V_{SSk_0}|^4}{16k^2_0\os^2(\ok{0}^2-4\os^2)^2+\lambda^2  |V_{SSk_0}|^4}+O(\lambda^3). \label{ir}
\eeq
Note that the $O(\lambda^0)$ term in the denominator has a zero, near which the $O(\lambda^2)$ term in the denominator dominates.  This is the usual $\Gamma^2$ term, which cuts off the pole reported in Ref.~\cite{Bilguun2025elastickinkmesonscatteringphi4}.

The probability $P$ of elastic scattering is $|R|^2$.  As there are no other imaginary contributions to the residue at this order with which this process may interfere, the leading contribution to this arising from the imaginary part of the residue is
\beq\label{DeltaP}
\Delta P=(\Im[R])^2=\frac{\lambda^4 |V_{SSk_0}|^8}{\left(16k^2_0\os^2(\ok{0}^2-4\os^2)^2+\lambda^2  |V_{SSk_0}|^4\right)^2}+O(\lambda^5).
\eeq
Notice that at the pole, the first term in the denominator vanishes and so we find a probability of reflection which is, in this leading pole approximation, equal to unity.  In other words, there is total reflection.  

Is this possible?  After all, the contribution to forward scattering from $\gamma_{2i}^{01}(k\sim k_0)$ follows from the same calculation and above, and so is given by the same residue Eq.~(\ref{ir}) which is equal to $-i$ at the pole.  As in backward scattering, one arrives at the forward scattering amplitude by integrating around this pole, closing the contour in the negative imaginary part of the complex plane.  This yields a forward scattering matrix element which, like the backward scattering matrix element, is equal to $-1$ at the Breit-Wigner peak.  However, the forward scattering amplitude, unlike the backward scattering amplitude, interferes coherently with the unscattered meson, which has a matrix element of $1$.  Altogether the forward scattering matrix element is then $1-1=0$, which is consistent with complete reflection at the peak.

How does this affect the pole in $R$ reported above?  Recall that the pole arises from the $D$ contribution in the last term of Eq.~(\ref{scat}).  Instead, we have found the following expression for $D$
\bea
D(k_0) &=& \frac{1}{8k_0}\int\frac{dk'_1dk'_2}{(2\pi)^2}\frac{(\omega_{k_1\p} + \omega_{k_2\p}) V_{-k_0 k_1\p k_2\p} V_{-k_0-k_1\p -k_2\p}}{\omega_{k_1\p} \omega_{k_2\p} \left(\omega_{k_0}^2 - \left(\omega_{k_1\p} + \omega_{k_2\p}\right)^2 + i \epsilon\right)} \label{dnuovo}\\ 
&&\hspace{-1cm}+ \frac{1}{8k_0}\pin{k^\prime}\frac{(\omega_{k'} + \omega_{S}) V_{-k_0 k'S} V_{-k_0-k'S}}{\omega_{k\p} \omega_{S} \left(\omega_{k_0}^2 - \left(\omega_{k\p} + \omega_{S}\right)^2 + i \epsilon\right)} +\frac{1}{4k_0}\frac{|V_{S Sk_0}|^2}{\omega_S \left(  \omega^2_{k_0}- 4 \omega_{S}^2\right)+i\lambda|V_{SSk_0}|^2}.
\nonumber
\eea
Now that $R$ has a finite imaginary part, one must remember \cite{evslin2023elastickinkmesonscattering}  that the probability is $|R|^2$.

\section{Example: The $\phi^4$ double-well} \label{modsez}

We now specialize our general expressions to the $\phi^4$ double-well model in $(1+1)$
dimensions, defined by the potential
\beq
V(\sqrt{\lambda}\phi)=\frac{1}{4}(\sqrt{\lambda}\phi)^2\left(\sqrt{\lambda}\phi-\sqrt{2}m\right)^2,
\eeq
so that the static kink solution is
\beq
\phi(x)=f(x)=\frac{m}{\sqrt{2\lambda}}\left(1+\tanh\!\left(\frac{mx}{2}\right)\right),
\eeq
with classical kink mass
\beq
Q_0=\frac{m^3}{3\lambda}.
\eeq

The translational zero mode, the (single) shape mode, and the continuum normal modes are respectively
\bea\label{modes}
\mathfrak{g}_B(x)&=&-\frac{\sqrt{3m}}{2\sqrt{2}}\sech^2\left(\frac{mx}{2}\right)\hsp
\mathfrak{g}_S(x)=\frac{3\sqrt{m}}{2\sqrt{3}}\tanh\left(\frac{mx}{2}\right)\sech\left(\frac{mx}{2}\right)
,\\
\mathfrak{g}_k(x)&=&\frac{2e^{-ikx}}{\omega_k\sqrt{m^2+4k^2}}
\left(
k^2-\frac{m^2}{2}
+\frac{3m^2}{4}\sech^2\left(\frac{mx}{2}\right)
-i\frac{3m}{2}k\tanh\left(\frac{mx}{2}\right)
\right),
\nonumber
\eea
with corresponding frequencies
\beq
\omega_B=0\hsp 
\omega_S=\frac{\sqrt{3}}{2}m\hsp
\omega_k=\sqrt{m^2+k^2}.
\eeq

The cubic vertex that mixes the two shape mode state into the one-meson continuum is~\cite{Bilguun2025elastickinkmesonscatteringphi4}
\beq
V_{SSk}
=
-i\pi\frac{3\sqrt{\lambda}}{\sqrt{2}}
\frac{k^2\omega_k(m^2-2k^2)}{m^3\sqrt{m^2+4k^2}}.
\label{eq:VSSk_phi4}
\eeq

In the $\phi^4$ model, there is only one real shape mode.   Ref.~\cite{Bilguun2025elastickinkmesonscatteringphi4} provided the corresponding amplitude from Eq.~(\ref{rdef}).  However, we have seen that the sum of bubble diagrams implies that we must change the $D_{SS}$ contribution. Eq.~(\ref{dnuovo}) implies that the new $D_{SS}$ term is 
\beq\label{D_SSP4}
D_{SS}(k_0)
=
\frac{9\pi^2\lambda k_0^{4} \omega_{k_0}^{2} \bigl(3m^2-2\omega_{k_0}^2\bigr)^2}
{8m^6 k_0 \omega_S(4\omega_{k_0}^2-3m^2)( \omega_{k_0}^2-4\omega_S^2)+i\lambda 9\pi^2 k_0^{4} \omega_{k_0}^{2} \bigl(3m^2-2\omega_{k_0}^2\bigr)^2}
 \csch^{2} \left(\frac{\pi k_0}{m}\right),
\eeq
while all the other contributions to the reflection amplitude, $A(k_0), B(k_0)$ and $C(k_0)$ are not touched since these terms do not have a pole corresponding to the twice-excited shape mode state. Note that the intermediate states $\ket{k'_1k'_2}$ and $\ket{Sk'}$ corresponding to the terms $D_{kk}(k_0)$ and $D_{Sk}(k_0)$ in the full $D(k_0)$ are also not changed, but have branch cuts \cite{Bilguun2025elastickinkmesonscatteringphi4}, therefore, 

\allowdisplaybreaks
\begin{align}
D(k_0) \label{dk0}
&= D_{SS}(k_0)+D_{Sk}(k_0)+D_{kk}(k_0) \nonumber \\
&=\frac{9\pi^2\lambda k_0^{4} \omega_{k_0}^{2} \bigl(3m^2-2\omega_{k_0}^2\bigr)^2}
{8m^6 k_0 \omega_S(4\omega_{k_0}^2-3m^2)( \omega_{k_0}^2-4\omega_S^2)+i\lambda 9\pi^2 k_0^{4} \omega_{k_0}^{2} \bigl(3m^2-2\omega_{k_0}^2\bigr)^2}
 \csch^{2} \left(\frac{\pi k_0}{m}\right) \nonumber \\
&\quad + \frac{27 \pi^2\lambda}{2^{14} k_0 m^3\omega^2_{k_0}\omega_S(4\omega^2_{k_0}-3m^2)} \nonumber \\
&\qquad \times \int\frac{dk'}{2\pi}\frac{(\omega_{k'} + \omega_S)[m^2 (17 m^4  +  68 m^2 k_{+}  +  32 k^2_{+}) - 16 k^2_{-} (3 m^2  +  4 k_{+})]^2}{\omega^3_{k'}(4\omega^2_{k'}-3m^2)(\omega^2_{k_0}-(\omega_{k'} + \omega_S)^2+i\epsilon)} \nonumber \\
&\qquad \times \sech\left(\frac{(k_0 - k')\pi}{m}\right)\sech\left(\frac{(k_0 + k')\pi}{m}\right) \nonumber \\
&\quad + F_{10}(k_0)\int\frac{dk'_1}{2\pi}F_{11}(k_0,k'_1)\left(Q_{-}(k_0,k'_1)-Q_{+}(k_0,k'_1)\right) \nonumber \\
&\quad + F_{21}(k_0)\int\frac{dk'_1 dk'_2}{(2\pi)^2}F_{22}(k_0,k'_1,k'_2).
\end{align}

The functions $F_{10}(k_0),F_{11}(k_0,k'_1)$ and $Q_{\mp}(k_0,k'_1)$ are:
\begin{equation}
    \begin{split}
        F_{10}(k_0) = & \frac{27m^2\pi\lambda}{2 \ok{0}^2 (4\omega^2_{k_0}-3m^2)}\csch\left(\frac{2k_0\pi}{m}\right), \\ 
        F_{11}(k_0,k'_1) = & \frac{k'_1 }{(4\omega^2_{k'_1}-3m^2)\omega^3_{k'_1}}, \\
        Q_{\mp}(k_0,k'_1)= & \frac{F_{\mp}(k_0,k'_1)P_{\mp}(k_0,k'_1)}{D_{\mp}(k_0,k'_1)},
    \end{split}
\end{equation}
with
\begin{equation}
    \begin{split}
        F_{\mp}(k_0,k'_1)=&(k_0\mp k'_1)(4(k^2_0\mp k_0k'_1+{k'_1}^{2})+3m^2), \\ 
        P_{\mp}(k_0,k'_1)=&(S_{\mp}(k_0,k'_1)+L_{\mp}(k_0,k'_1))(\omega_{k'_1} + \omega_{k_0\mp k'_1}), \\ 
        S_{\mp}(k_0,k'_1)=& 8k^2_0(k_0\mp k'_1)^2{k'_1}^2-4(k^2_0\mp k_0k'_1+{k'_1}^2)^2m^2, \\
        L_{\mp}(k_0,k'_1)=& -5(k^2_0\mp k_0k'_1 +{k'_1}^2)m^4-m^6, \\
        D_{\mp}(k_0,k'_1)=&({4\omega^2_{{k_0\mp k'_1}}} -3 m^2)\omega^3_{{k_0\mp k'_1}}(\omega^2_{k_0} - (\omega_{k'_1} + \omega_{{k_0\mp k'_1}})^2 + i\epsilon).
    \end{split}
\end{equation}

The functions $F_{21}(k_0)$ and $F_{22}(k_0,k'_1,k'_2)$ in the $D_{kk}(k_0)$ are:

\begin{equation}
    \begin{split}
        F_{21}(k_0)=&\frac{-9\pi^2\lambda}{4k_0 \ok{0}^2 (4\omega^2_{k_0}-3m^2)},\\
        F_{22}(k_0,k'_1,k'_2)=&\frac{[f(k_0,k'_1,k'_2)]^2 (\omega_{k'_1}+\omega_{k'_2})}{D_0(k'_1,k'_2)D_1(k'_1)D_2(k'_2)D_3(k_0,k'_1,k'_2)}\\
        & \times\csch\left(\frac{(k_0 - k'_1 - k'_2)\pi}{m}\right)\csch\left(\frac{(k_0 + k'_1 + k'_2)\pi}{m}\right).
    \end{split}
\end{equation}
The energy denominators are 
 \begin{equation}
     \begin{split}
         D_0(k'_1,k'_2)=& \omega^3_{k'_1}\omega^3_{k'_2}, \quad
         D_{1,2}(k'_{1,2})= (4\omega^2_{k'_{1,2}}-3m^2), \quad 
         D_3(k_0,k'_1,k'_2)=(\omega^2_{k_0}-(\omega_{k'_1}+\omega_{k'_2})^2+i\epsilon).
     \end{split}
 \end{equation}
The numerator function $f(k_0,k'_1,k'_2)$  and its coefficients are
\begin{equation}
    \begin{split}
        &f(k_0,k'_1,k'_2)=3a^3 - 3a^2s - a(3s^2 - 4p + 8sm^2 + 5m^4) + 3s^3 - 12sp - 8pm^2 - 5sm^4 - 2m^6, \\ 
        &a=k^2_0, \quad s={k'_1}^2+{k'_2}^2, \quad p={k'_1}^2{k'_2}^2, \quad k_{\pm}=k^2_0\pm {k'}^2.
    \end{split}
\end{equation}
The resonance occurs when the incoming meson is on shell with twice the shape mode energy,
$\omega_{k_0}=2\omega_S$, which gives
\beq
k_0=\sqrt2m,
\qquad
\omega_S=\frac{\sqrt{3}}{2}m.
\eeq
The twice-excited shape mode has energy $2\omega_{S}=\sqrt{3}m$, which is above the one-meson energy but below the two-meson energy; therefore, near the $\ket{SS}$ state, the single-meson-bubble energy dominates. 
The decay width of the twice-excited shape mode state is, 
\begin{equation}\label{Gamma}
    \Gamma_{\phi^4}=\frac{9\pi^2}{\sqrt{2}}\csch^2(\sqrt{2}\pi)\frac{\lambda}{m},
\end{equation}
which also agrees with the decay width found in Ref.~\cite{Evslin_2022}. 

\begin{figure}[htbp]
\centering
\includegraphics[width=.6\textwidth]{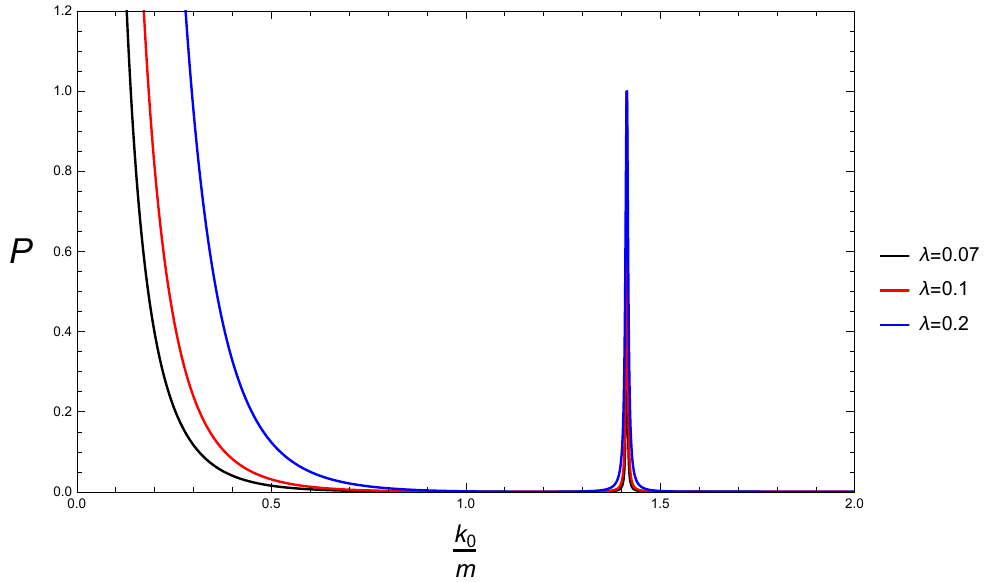}
\includegraphics[width=.48\textwidth]{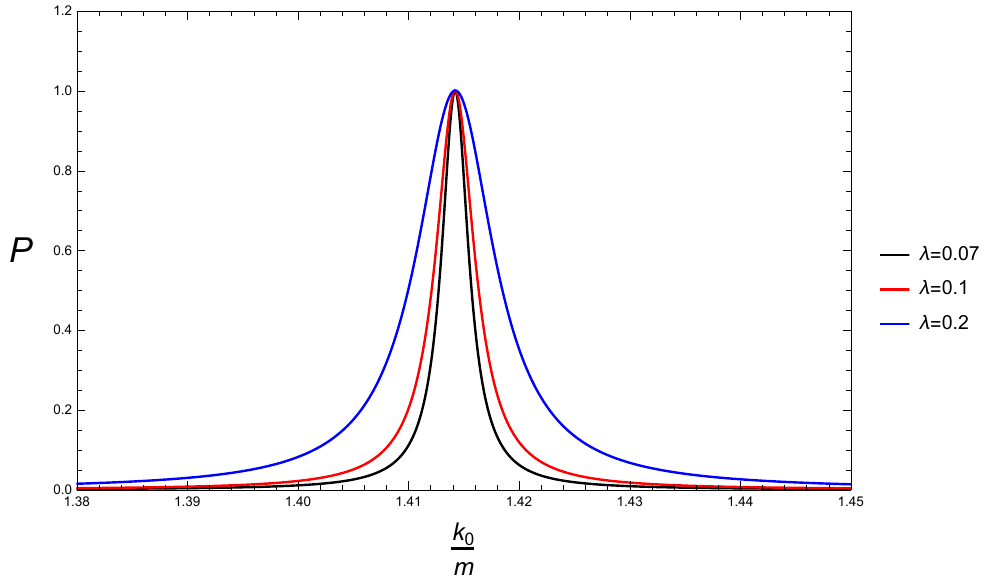}
\includegraphics[width=.48\textwidth]{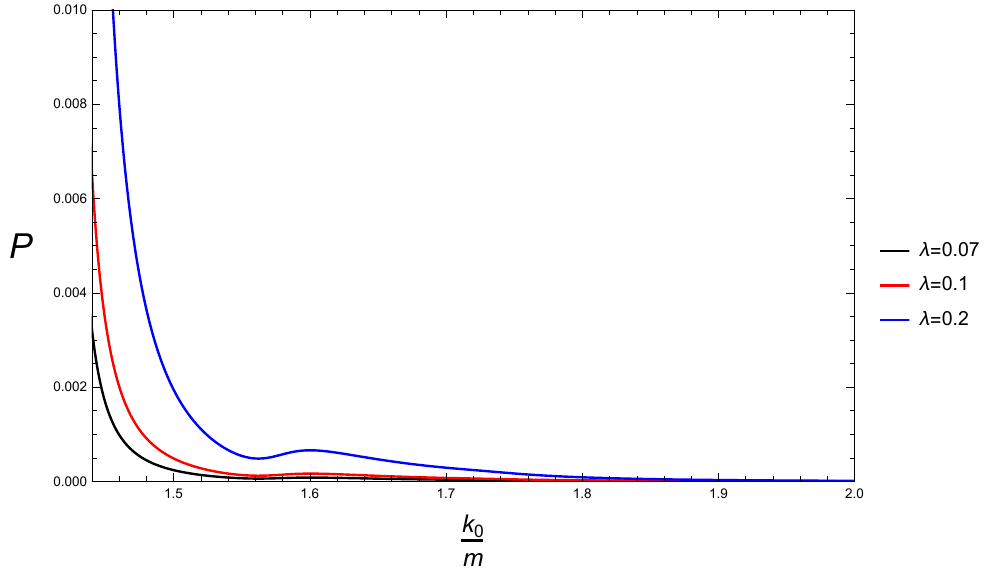}
\caption{The leading reflection probability as a function of the initial momentum $k_0$.  The bottom panels are respectively zoomed near the resonance in $D_{SS}$ and near the branch point in $D_{Sk}$, which is the threshold for an intermediate state with a once-excited shape mode and a virtual meson.  Subleading corrections will make the total probability less than unity.}
\label{ph4ImR1}
\end{figure}

Fig.~\ref{ph4ImR1} shows the probability $P(k_0)$ with the enhanced $D_{SS}$ contribution Eq.~\eqref{D_SSP4}, where the probability  $P(k_0)$ and reflection amplitude $R(k_0)$ are 
\begin{equation}\label{Pk0}
    P(k_0)=(\Re[R(k_0)])^2+(\Im[R(k_0)])^2,\quad R(k_0)=\lambda(A(k_0)+B(k_0)+C(k_0)+D(k_0)).
\end{equation}

\section{Remarks}
\label{sec:remarks}

In the classical $\phi^4$ double-well field theory, at linear level, radiation passes through the kink without interacting\footnote{At the nonlinear level, kink-radiation interactions have a rich phenomenology, exhibiting phenomena such as the negative radiation pressure of Refs.~\cite{neg1,neg2}, which has since been extended to other classical solutions \cite{nega1,nega2}.}.  In quantum field theory, this is not the case~\cite{evslin2023elastickinkmesonscattering}.  A meson striking a kink has a finite probability of elastic scattering, and so the scattered power is linear in the power of the incoming radiation when both are small.  In the present paper we have seen that this elastic scattering process, which is only allowed in the quantum theory, opens the window to new observables: the energies and lifetimes of the kink's unstable excitations.  Strictly speaking, these new observables also only exist in the quantum theory.  However, at tree-level the lifetime of the unstable state can be used, and indeed has been used \cite{Evslin_2022}, to calculate the decay rate that results from nonlinear dynamics in the classical field theory \cite{mm}.  In other words, a resonance in a linear process in quantum field theory tells us about the nonlinear dynamics of the resonant state in the classical field theory, and also about its quantum corrections.

In this work we obtained the leading order lifetime of the twice-excited shape mode state by summing the bubble contributions that dominate elastic
kink-meson scattering in the vicinity of the two shape mode threshold. At leading order in the coupling, the reflection
amplitude contains an explicit pole associated with an intermediate kink state carrying two
shape mode quanta \cite{Bilguun2025elastickinkmesonscatteringphi4}. In the near-resonant regime,
successive bubble insertions are not parametrically suppressed and must be summed. The result is
a Dyson-type dressing of the two shape mode denominator governed by the self-energy, transforming the pole into a Breit-Wigner peak with width fixed by Eq.~\eqref{Gamma}.


Several extensions suggest themselves. First, away from the narrow-resonance regime, a quantitative
description of the full reflection probability requires combining the resummed resonant
contribution with the smooth background arising from the remaining one-loop terms
\cite{Bilguun2025elastickinkmesonscatteringphi4}. Second, at higher orders additional intermediate
states in the kink sector contribute to the self-energy, including multi-meson channels, and new
thresholds will generically distort the line shape. Third, in models with multiple bound modes or
nontrivial reflection data, coupled-channel dynamics can produce richer resonant structure; it
would be interesting to apply the same resummation strategy in those settings. Finally, relaxing
the assumption of equal meson masses in the two vacua introduces one-loop acceleration effects
\cite{Weigel_2019} and ultimately calls for a time-dependent treatment along the lines of
Ref.~\cite{evslin2024perturbativeapproachtimedependentquantumv}.


More ambitiously, at large $N$ baryons become solitons in the meson field.  Although the Skyrme model describing these solitons at leading order in the large $N$ expansion is not renormalizable, nonetheless the scattering calculation can also be applied to that case, yielding the spectrum of nucleon excitations in the large $N$ limit together with their lifetimes.  Of course, this problem would be numerically challenging, to say the least, because of the need to understand and numerically manipulate the linearized perturbations about the Skyrmion~\cite{spert1,spert2}.
 
\section* {Acknowledgement}

\noindent
BB is supported by the ANSO scholarship CAS.  This work was supported by the Higher Education and Science Committee of the Republic of Armenia (Research Project No. 24RL-1C047).

\bibliographystyle{elsarticle-num} 
\bibliography{ref} 

\end{document}